\documentclass[prd,aps,superscriptaddress,onecolumn,preprintnumbers,floatfix,nofootinbib]{revtex4-1}

\usepackage[T1]{fontenc}
\usepackage[english]{babel}

\usepackage{graphicx}	
\usepackage{amsmath}	
\usepackage{amssymb}	
\usepackage{amsfonts}
\usepackage{amsbsy}
\usepackage{amstext}
\usepackage{color}
\usepackage{rotating}
\usepackage{multirow}
\usepackage{dcolumn}
\usepackage{bm}
\usepackage{newtxtext,newtxmath}
\usepackage{siunitx}

\newcommand{\be}{\begin{equation}}
\newcommand{\ee}{\end{equation}}
\newcommand{\ba}{\begin{eqnarray}}
\newcommand{\ea}{\end{eqnarray}}
\DeclareSIUnit\parsec{pc}

\begin{document}

\title{J-PAS: forecasts on interacting vacuum energy models}

\author{V.~Salzano}\email{vincenzo.salzano@usz.edu.pl}
\affiliation{Institute of Physics, University of Szczecin, Wielkopolska 15, 70-451 Szczecin, Poland}
\author{C.~Pigozzo}
\affiliation{Instituto de F\'{\i}sica, Universidade Federal da Bahia, 40210-340, Salvador, BA, Brazil}
\author{M.~Benetti}
\affiliation{Dipartimento di Fisica ``E. Pancini'', Universit\`a di Napoli ``Federico II'', Via Cinthia, I-80126, Napoli, Italy}
\affiliation{Istituto Nazionale di Fisica Nucleare (INFN), sez. di Napoli, Via Cinthia 9, I-80126 Napoli, Italy}
\author{H.~A.~Borges}
\affiliation{Instituto de F\'{\i}sica, Universidade Federal da Bahia, 40210-340, Salvador, BA, Brazil}
\author{R.~von Marttens}
\affiliation{Departamento de Astronomia, Observat\'orio Nacional, Rio de Janeiro, 20921-400, RJ, Brazil}
\author{S.~Carneiro}
\affiliation{Instituto de F\'{\i}sica, Universidade Federal da Bahia, 40210-340, Salvador, BA, Brazil}
\author{J.~S.~Alcaniz}
\affiliation{Departamento de Astronomia, Observat\'orio Nacional, Rio de Janeiro, 20921-400, RJ, Brazil}
\author{J.~C.~Fabris}
\affiliation{N\'ucleo Cosmo-ufes \& Departamento de F\'{\i}sica, Universidade Federal do Esp\'{\i}rito Santo, 29075-910, Vit\'oria, ES, Brazil}
\author{S.~Tsujikawa}
\affiliation{Department of Physics, Waseda University, 3-4-1 Okubo, Shinjuku, Tokyo 169-8555, Japan}
\author{N.~Ben\'itez}
\affiliation{Instituto de Astrof\'isica de Andaluc\'ia (CSIC), Glorieta de la Astronom\'ia s/n, Granada, 18008, Spain}
\author{S.~Bonoli}
\affiliation{Centro de Estudios de F\'{\i}sica del Cosmos de Arag\'on (CEFCA), Plaza San Juan, 1, E-44001, Teruel, Spain}
\affiliation{Donostia International Physics Center (DIPC), Manuel Lardizabal Ibilbidea, 4, San Sebasti\'an, Spain}
\affiliation{Ikerbasque, Basque Foundation for Science, E-48013 Bilbao, Spain}
\author{A.~J.~Cenarro}
\affiliation{Centro de Estudios de F\'{\i}sica del Cosmos de Arag\'on (CEFCA), Unidad Asociada al CSIC, Plaza de San Juan, 1, E-44001, Teruel, Spain}
\author{D.~Crist\'obal-Hornillos}
\affiliation{Centro de Estudios de F\'{\i}sica del Cosmos de Arag\'on (CEFCA), Unidad Asociada al CSIC, Plaza de San Juan, 1, E-44001, Teruel, Spain}
\author{R.~A.~Dupke}
\affiliation{Departamento de Astronomia, Observat\'orio Nacional, Rio de Janeiro, 20921-400, RJ, Brazil}
\affiliation{Department of Astronomy, University of Michigan, 311 West Hall, 1085 South University Ave., Ann Arbor, USA}
\author{A.~Ederoclite}
\affiliation{Departamento de Astronomia, Instituto de Astronomia, Geof\'isica e Ci\^encias Atmosf\'ericas, Universidade de S\~ao Paulo, R. do Mat\~ao 1226, S\~ao Paulo, SP 05508-090, Brazil}
\author{C.~L\'opez-Sanjuan}
\affiliation{Centro de Estudios de F\'{\i}sica del Cosmos de Arag\'on (CEFCA), Unidad Asociada al CSIC, Plaza de San Juan, 1, E-44001, Teruel, Spain}
\author{A.~Mar\'in-Franch}
\affiliation{Centro de Estudios de F\'{\i}sica del Cosmos de Arag\'on (CEFCA), Unidad Asociada al CSIC, Plaza de San Juan, 1, E-44001, Teruel, Spain}
\author{V.~Marra}
\affiliation{N\'{u}cleo de Astrof\'{\i}sica e Cosmologia \& Departamento de F\'{\i}sica, Universidade Federal do Espírito Santo, 29075-910, Vit\'{o}ria, ES, Brazil}
\affiliation{INAF -- Osservatorio Astronomico di Trieste, via Tiepolo 11, 34131 Trieste, Italy}
\affiliation{IFPU -- Institute for Fundamental Physics of the Universe, via Beirut 2, 34151, Trieste, Italy}
\author{M.~Moles}
\affiliation{Centro de Estudios de F\'{\i}sica del Cosmos de Arag\'on (CEFCA), Unidad Asociada al CSIC, Plaza de San Juan, 1, E-44001, Teruel, Spain}
\author{C.~Mendes de Oliveira}
\affiliation{Departamento de Astronomia, Instituto de Astronomia, Geof\'isica e Ci\^encias Atmosf\'ericas, Universidade de S\~ao Paulo, R. do Mat\~ao 1226, S\~ao Paulo, SP 05508-090, Brazil}
\author{L.~Sodr\'e Jr}
\affiliation{Departamento de Astronomia, Instituto de Astronomia, Geof\'isica e Ci\^encias Atmosf\'ericas, Universidade de S\~ao Paulo, R. do Mat\~ao 1226, S\~ao Paulo, SP 05508-090, Brazil}
\author{K.~Taylor}
\affiliation{Instruments4, 4121 Pembury Place, La Ca\~{n}ada-Flintridge, Ca 91011, USA}
\author{J.~Varela}
\affiliation{Centro de Estudios de F\'{\i}sica del Cosmos de Arag\'on (CEFCA), Unidad Asociada al CSIC, Plaza de San Juan, 1, E-44001, Teruel, Spain}
\author{H.~V\'azquez Rami\'o}
\affiliation{Centro de Estudios de F\'{\i}sica del Cosmos de Arag\'on (CEFCA), Unidad Asociada al CSIC, Plaza de San Juan, 1, E-44001, Teruel, Spain}

\date{\today}

\begin{abstract}
The next generation of galaxy surveys will allow us to test some fundamental aspects of the standard cosmological model, including the assumption of a minimal coupling between the components of the dark sector. In this paper, we present the Javalambre Physics of the
Accelerated Universe Astrophysical Survey (J-PAS) forecasts on a class of unified models where cold dark matter interacts with a vacuum energy, considering future observations of baryon acoustic oscillations, redshift-space distortions, and the matter power spectrum. After providing a general framework to study the background and linear perturbations, we focus on a concrete interacting model without momentum exchange by taking into account the contribution of baryons. We compare the J-PAS results with those expected for DESI and Euclid surveys and show that J-PAS is competitive to them, especially at low redshifts. Indeed, the predicted errors for the interaction parameter, which measures the departure from a $\Lambda$CDM model, can be comparable to the actual errors derived from the current data of cosmic microwave background temperature anisotropies.
\end{abstract}


\maketitle

\section{Introduction}
\label{sec:introduction}

The quest to uncover the origin of the underlying cause of late-time
cosmic acceleration is one of the most important topics
in cosmology \citep{CST,Clifton,Joyce,Kase:2018aps}.
In the context of the $\Lambda$-Cold-Dark-Matter ($\Lambda$CDM)
model \citep{Peebles1,Peebles2},
gravity is described by Einstein's General Relativity in the presence of the cosmological constant
$\Lambda$ and non-relativistic dark matter. Although the $\Lambda$CDM model is
treated as a standard cosmological paradigm,  the origins of the dark sector are largely unknown.
Moreover, the recent observations have shown that, in the $\Lambda$CDM model,
there are tensions for today's Hubble constant
$H_0$ between the measurements of
Cosmic Microwave Background (CMB) temperature anisotropies \citep{Aghanim:2018eyx}
and its direct measurements at low redshifts \citep{Riess:2019cxk,Freedman:2019jwv}.

Additionally, a basic hypothesis of standard cosmology is that non-relativistic matter,
which forms galaxies and clusters, obeys the continuity equation by its own.
This conserved property holds if dark energy, the mechanism behind the late-time
acceleration, is a non-interacting vacuum energy $\Lambda$.
If CDM interacts with $\Lambda$, the vacuum energy can be
time-dependent \citep{Ozer:1985wr,Freese:1986dd,Berman:1991zz,Pavon:1991uc,
Carvalho:1991ut,Shapiro:2000dz,PLB1,PLB2,winfried,wands2,iModel8,micol,vonMarttens:2018iav}.
Although the presence of $\Lambda$ does not give rise to any dynamical degrees
of freedom, the modified background evolution induced by the coupling with CDM
changes the cosmic distance to sources.
Hence the interacting vacuum energy scenario can be observationally
distinguished from the uncoupled $\Lambda$CDM model \citep{Xu:2011qv}.

In interacting vacuum energy models where the coupling is phenomenologically
chosen to realize a desired background evolution with late-time cosmic acceleration,
it is non-trivial to deal with the dynamics of inhomogeneous perturbations in a consistent way.
\citep{wands2} addressed this issue by considering the vacuum-energy perturbation
$\delta \Lambda$ induced by the energy or momentum exchanges with CDM.
Reflecting the fact that the four velocity of vacuum energy is undetermined,
we need to make an assumption that $\delta \Lambda$ is related to
other perturbations. One way is to assume that $\Lambda$ depends on
the CDM density $\rho_c$ \citep{wands2}, in which case $\delta \Lambda$ has a relation
with the CDM perturbation $\delta \rho_c$.
The other possible way is to consider the case in which either
energy or momentum exchange between $\Lambda$ and
CDM is absent \citep{wands}.
In this paper we will provide a general theoretical
framework of interacting vacuum energy with CDM by taking
into account baryons, but for the purpose of the J-PAS forecast,
we will choose a concrete model in which the momentum
exchange is absent.

Without restricting the origin of late-time cosmic acceleration
to the cosmological constant, there are also other phenomenological
approaches to dynamical dark energy models coupled to CDM
in which the dark energy equation of
state $w_{\rm DE}$ deviates from
$-1$ \citep{Dalal:2001dt,Zimdahl:2001ar,iModel1,Chimento:2003iea,peebles,zero,ademir,Amendola:2006dg, Wei:2006ut,setare,Ira,super,Guo:2007zk,bertolami,Valiviita:2008iv,Gavela:2009cy,
iModel2,iModel3,iModel4,maartens,Wang:2016lxa}.
Numerous observational data including CMB, supernovae type Ia (SNeIa), and Baryon
Acoustic Oscillations (BAO) have been used to place constraints on such
interacting models  (see e.g., \citep{cid} and references therein).
In particular, it was shown that the presence of couplings between dark energy and
CDM can alleviate the observational tension of $H_0$ present in the $\Lambda$CDM
model \citep{Kumar:2016zpg,iModel11,DiValentino:2017iww,An:2017crg,Yang:2018euj,
Pan:2019gop,DiValentino:2019ffd,Yang:2019uog,DiValentino:2019jae,Vagnozzi:2019kvw}.
There are also dynamical interacting models of dark energy with some
concrete Lagrangians of scalar fields or vector
fields \citep{Wette,Luca,Pourtsidou:2013nha,Boehmer:2015kta,Boehmer:2015sha,Skordis:2015yra,
Koivisto:2015qua,Pourtsidou:2016ico,Dutta:2017kch,Linton,Kase:2019veo,Kase:2019mox,
Chamings:2019kcl,Amendola:2020ldb,Kase:2020hst,DeFelice:2020icf},
in which case it is possible to reduce observational
tensions of $\sigma_8$.
In interacting models of dark perfect fluids with a momentum exchange,
it was recently shown that the observational tension of the $\Lambda$CDM model
can be also eased \citep{Asghari:2019qld,Jimenez:2020ysu,FiguerueloJPAS}.

To place observational constraints on interacting models of dark energy,
the next generation of galaxy surveys, such as DESI~\citep{Aghamousa:2016zmz},
{Euclid}~\citep{Euclid}, SKA~\citep{ska} and Javalambre Physics of the
Accelerated Universe Astrophysical Survey (J-PAS)~\citep{Benitez:2014ibt,mini},
will play major roles. In these models, not only the background evolution
but also the dynamics of cosmological perturbations is modified in comparison
to the $\Lambda$CDM. With high-precision observational data, there will be a
chance of detecting the signature of dark sector interactions,
in spite of dark degeneracy mentioned in
Refs.~\citep{Degeneracy1,Degeneracy2,Degeneracy3,CB,vonMarttens:2019ixw,vonMarttens:2020apn}.
In this paper, we consider the simple interacting vacuum energy model mentioned
above and discuss expected cosmological impacts extracted from J-PAS \citep{Benitez:2014ibt,mini}
measurements. It is straightforward to extend the analysis to more general dynamical coupled dark energy
models, but we will leave it for future work due to their complexities.

Our purpose in this paper is twofold. First, we study the implications of J-PAS
on a concrete interacting model characterized by a coupling
constant $\alpha$ (defined in Sec.~\ref{concretesec}).
Second, we compare the J-PAS forecast results with those expected from the
DESI~\citep{Aghamousa:2016zmz} and { Euclid}~\citep{Euclid} surveys.
For two configurations of area, i.e., 4000 ${\rm{deg^2}}$ and 8500 ${\rm{deg^2}}$,
we consider the J-PAS information on BAO, Redshift-Space Distortions (RSD),
and the full matter power spectrum (PS), and perform a Fisher matrix forecast on
the parameter $\alpha$, which also quantifies a possible deviation from the $\Lambda$CDM cosmology ($\alpha = 0$). In order to improve the accuracy of estimates, we also perform a multi-tracer analysis and compare all J-PAS forecasts  with those expected by the DESI and Euclid surveys. We find that J-PAS will be able to measure the interaction parameter with good sensitivity,
and will provide the best constraints on interaction in the redshift range $z \lesssim 0.5$,
thanks to the large number
of emission line galaxies detectable in that redshift range.

The outline of this paper is as follows.
Section \ref{jpassec} briefly introduces the main characteristics of the J-PAS.
General aspects of interacting vacuum energy including the background
equations of motion are discussed in Sec.~\ref{genesec}.
In Sec.~\ref{persec}, we derive the perturbation equations in a gauge-ready form
and discuss different choices of gauges.
In Sec.~\ref{concretesec}, we present a concrete model of interacting vacuum
energy and study the background and perturbation dynamics taking
into account baryon perturbations.
Sec.~\ref{Fishersec} presents our Fisher matrix analysis,
whereas Sec.~\ref{resultsec} discusses our main results, including a comparison
with forecasts for the DESI and Euclid surveys.
We summarize our main conclusions in Sec.~\ref{concludesec}.

\section{J-PAS}
\label{jpassec}

The J-PAS~\citep{Benitez:2014ibt,mini} is a spectro-photometric,  ground-based imaging survey
that will observe 8500 ${\rm{deg^2}}$  of the northern sky.
The survey will be conducted  at the Observatorio Astrof\'{\i}sico de Javalambre,
a site on top of Pico del Buitre, a summit about  $\sim 2000$ m high above sea
 level at the Sierra de Javalambre (Spain).

The survey will be carried out in a seven-year observing program at a dedicated 2.5 m telescope,
the Javalambre Survey Telescope (JST/T250), equipped with a 1.2 Gigapixel camera (JPCam)
with a large field of view of 4.2 ${\rm{deg^2}}$,
and will incorporate a 54 narrow- and 4 broad-band filter set covering the optical range.
J-PAS will measure positions and redshifts for dozens of millions of Luminous Red Galaxies (LRG),
Emission Line Galaxies (ELG) and millions of Quasars (QSO), with an expected photometric redshift precision
of $\sigma(z) = 0.003(1 + z)$, where $z$ is the redshift.
Moreover, given its high photo-$z$ precision, J-PAS will detect nearly a hundred thousand clusters
of galaxies up to the redshift $z \sim 1$ and several times more groups of galaxies.
They can be used to improve the constraints on cosmological parameters not just through
the mass function but also through high-precision definition of the cosmic web.
For more details on J-PAS, we refer the reader to \citep{Benitez:2014ibt,mini}.

\section{General Aspects of Interacting Vacuum Energy}
\label{genesec}

We consider cold dark matter interacting with
a vacuum energy $\Lambda$ \citep{Ozer:1985wr,Freese:1986dd,Berman:1991zz,Pavon:1991uc,
Carvalho:1991ut,Shapiro:2000dz,PLB1,PLB2,micol,winfried,wands2,iModel8}.
The CDM is described by a pressureless perfect fluid
given by the energy-momentum tensor
\be
T^{\mu \nu}_{c} = \rho_c u^{\mu} u^{\nu}\,,
\label{Tmunu1}
\ee
where $\rho_c$ is the CDM density, and $u^{\mu}$
is the four velocity obeying the relation
$u_{\mu} u^{\mu}=-1$.
The energy-momentum tensor of vacuum is
given by
\be
T^{\mu \nu}_{\Lambda} =  -\Lambda g^{\mu \nu}\,,
\label{Tmunu2}
\ee
where $g^{\mu \nu}$ is the metric tensor.
The vacuum energy density $\hat{\rho}$ and
pressure $\hat{P}$ satisfy $\hat{\rho}=-\hat{P}=\Lambda$, i.e.,
$\hat{\rho}+\hat{P}=0$, so its four velocity
$\hat{u}_{\mu}$ is undefined.
Unlike CDM, the vacuum energy does not act
as a dynamical degree of freedom.
Taking the covariant derivative of Eq.~(\ref{Tmunu2}),
it follows that
\be
{T^{\mu \nu}_{\Lambda}}_{;\nu}=-Q^{\mu}\,,
\label{Tcon1}
\ee
where
\be
Q^{\mu}=\Lambda_{,\nu}g^{\mu \nu}\,.
\label{Qmu1}
\ee
The semicolon and colon represent the covariant and
partial derivatives, respectively.
If there is the energy or momentum transfer between
CDM and vacuum energy, we have $Q^{\mu} \neq 0$ and
hence the  vacuum energy becomes inhomogeneous
in spacetime \citep{wands2}.

We assume that the total energy-momentum tensor
$T^{\mu \nu}=T_{c}^{\mu \nu}+T_{\Lambda}^{\mu \nu}$
satisfies the continuity equation ${T^{\mu \nu}}_{;\nu}=0$.
On using Eq.~(\ref{Tcon1}), the CDM sector obeys
\be
{T^{\mu \nu}_{c}}_{;\nu}=+Q^{\mu}\,.
\label{Tcon2}
\ee
We decompose $Q^{\mu}$ in the form
\be
Q^{\mu}=Qu^{\mu}+q^{\mu}\,,
\label{Qmu2}
\ee
where $q^{\mu}$ satisfies
\be
q^{\mu} u_{\mu}=0\,.
\label{Qcon}
\ee
Substituting Eq.~(\ref{Qmu2}) into Eq.~(\ref{Qmu1}) and
multiplying it by $u_{\mu}$, it follows that
\be
Q=-\Lambda_{,\nu} u^{\nu}\,.
\label{Qmu3}
\ee
On using Eqs.~(\ref{Qmu1}), (\ref{Qmu2}),
and (\ref{Qmu3}), we obtain
\be
q^{\mu}=\Lambda_{,\nu} \left(
u^{\mu} u^{\nu}+g^{\mu \nu} \right)\,.
\label{barQ}
\ee
The quantities $Q$ and $q^{\mu}$ represent the energy
transfer and momentum transfer, respectively.
Indeed, the energy exchange between two dark components is explicit
by taking the products of $u_{\mu}$ in Eqs.~(\ref{Tcon1})
and (\ref{Tcon2}) \citep{Amendola:2020ldb,Kase:2020hst},
such that
\begin{align}
u_{\mu} {T^{\mu \nu}_{\Lambda}}_{;\nu}&=+Q\, , \label{enecon1} \\
u_{\mu} {T^{\mu \nu}_{c}}_{;\nu}&=-Q\, . \label{enecon2}
\end{align}
On the right hand-side of these equations,
the terms associated with the momentum transfer
vanish because of the condition (\ref{Qcon}).

Let us consider the spatially flat
Friedmann-Lama\^itre-Robertson-Walker (FLRW)
spacetime given by the line element
\be
{\rm d}s^2=-{\rm d}t^2+a^2(t) \delta_{ij}
{\rm d}x^i {\rm d}x^j\,,
\label{lineb}
\ee
where $a$ is the scale factor which depends on
the cosmic time $t$.
On this background, the CDM four velocity can be chosen
as $u^{\mu}=(1,0,0,0)$ and hence the momentum transfer (\ref{barQ})
vanishes. Then, from Eqs.~(\ref{enecon1}) and (\ref{enecon2}),
it follows that
\begin{align}
\dot{\Lambda}&=-Q\,,
\label{cont1} \\
\dot{\rho}_c + 3H\rho_c&=Q\,,
\label{cont2}
\end{align}
where a dot represents a derivative with respect to $t$,
and $H \equiv \dot{a}/a$ is the Hubble expansion rate.
In the gravity sector we consider general relativity with
the Einstein equation
\be
G^{\mu \nu}=8 \pi G \left( T_c^{\mu \nu}
+T_\Lambda^{\mu \nu} \right)\,,
\label{Ein}
\ee
where $G^{\mu \nu}$ is the Einstein tensor, and $G$
is the gravitational constant.
We neglect the contribution of baryons to the right
hand-side of Eq.~(\ref{Ein}), but we will include it
in Secs.~\ref{basec}-\ref{growsec}.
The background equations of motion following from
Eq.~(\ref{Ein}) are
\begin{align}
3H^2 &= 8\pi G \left( \rho_c+\Lambda \right)\,,
\label{back1}\\
\dot{H} &= -4\pi G \rho_c\,,
\label{back2}
\end{align}
where Eq.~(\ref{back2}) can be also derived by taking the time
derivative of Eq.~(\ref{back1}) and employing
Eqs.~(\ref{cont1}) and (\ref{cont2}).

\section{Cosmological perturbations}
\label{persec}

Let us consider scalar perturbations on the flat FLRW background (\ref{lineb}).
The perturbed line element, which contains four metric perturbations
$A$, $B$, $\psi$, and $E$, is given by \citep{Bardeen}
\begin{align}\label{permet}
{\rm d}s^2=-\left( 1+2A \right) {\rm d}t^2+2a \partial_i B
{\rm d}t {\rm d}x^i + a^2\left[ \left( 1-2\psi \right) \delta_{ij}
+2\partial_i \partial_j E \right] {\rm d}x^i {\rm d}x^j\,,
\end{align}
where we used the notation $\partial_i \equiv \partial/\partial x^i$.
Since we are interested in the evolution of linear perturbations,
we neglect the terms higher than the first order
in the following discussion.

For CDM, the energy density is decomposed into the background and
perturbed parts, as $\rho_c=\bar{\rho}_c+\delta \rho_c$.
The CDM four velocity is defined by
$u^{\mu}={\rm d}x^{\mu}/{\rm d}\tau$, where $\tau$ is
the proper time satisfying ${\rm d}\tau=(1+A){\rm d}t$ for
the perturbed line element (\ref{permet}).
We express the spatial component of $u^{\mu}$ as
$u^i=a^{-1}\partial^i v$, where $v$ is the scalar velocity potential.
Then, the CDM four velocities with upper and lower indices
are expressed as
\be
u^{\mu}=\left( 1-A, a^{-1} \partial^i v \right)\,,\qquad
u_{\mu}=\left( -1-A, \partial_i \theta \right)\,,
\label{umu}
\ee
where
\be
\theta \equiv a \left( v+B \right)\,.
\ee
The four velocity of vacuum energy, $\hat{u}_{\mu}$, can be defined by
using the energy-momentum flow (\ref{Qmu1}), such that
\be
\hat{u}_{\mu}=-\frac{\Lambda_{,\mu}}
{|\Lambda_{,\nu} \Lambda^{,\nu}|^{1/2}}\,,
\label{hatmu}
\ee
which obeys the relation $\hat{u}_{\mu} \hat{u}^{\mu} =-1$
for the time-like flow ($\Lambda_{,\nu} \Lambda^{,\nu}<0$).
By introducing the vacuum energy velocity potentials
$\hat{v}$ and $\hat{\theta}$, as
$\hat{u}^i=a^{-1}\partial^i \hat{v}$ and
$\hat{\theta}=a(\hat{v}+B)$,
we can express $\hat{u}^{\mu}$ in a fashion
analogous to Eq.~(\ref{umu}):
\be
\hat{u}^{\mu}=\left( 1-A, a^{-1} \partial^i \hat{v} \right)\,,\qquad
\hat{u}_{\mu}=\left( -1-A, \partial_i \hat{\theta} \right)\,.
\label{umu2}
\ee
The vacuum energy has a perturbation $\delta \Lambda$,
which is induced by the interaction with CDM.
{}From Eq.~(\ref{hatmu}) the spatial component of $\hat{u}_{\mu}$ is
$\hat{u}_{i}=-\partial_i \delta \Lambda/\dot{\Lambda}$,
so the comparison with the second of Eq.~(\ref{umu2}) gives
\be
\delta \Lambda=-\dot{\Lambda} \hat{\theta}
=Q\hat{\theta}\,,
\label{delLam}
\ee
where we used Eq.~(\ref{cont1}).
This shows that $\delta \Lambda$ is directly related to
the velocity potential of vacuum energy.

\subsection{Perturbation equations}

We split the scalar quantity $Q$ into the background and
perturbed parts, as $Q=\bar{Q}+\delta Q$.
In the following, we omit a bar from the background quantities.
Then, the linear perturbation equations following from
Eqs.~(\ref{enecon1}) and (\ref{enecon2})
are given, respectively, by \citep{wands2}
\begin{align}
&\dot{\delta \Lambda}=-\delta Q-QA\, ,
\label{pereq1} \\
&\dot{\delta \rho}_c+3H \delta \rho_c-3\rho_c \dot{\psi}
+\rho_c \frac{\nabla^2}{a^2} \left( \theta+a^2 \dot{E}
-aB \right)=\delta Q+Q A\, ,
\label{pereq2}
\end{align}
where $\nabla^2$ is the three-dimensional Laplacian.
The perturbation of spatial components of $Q^\mu$
in Eq.~(\ref{Qmu2}) is expressed as
\be
\delta Q^i=Q u^i+\delta q^i
=\partial^i \left( a^{-1} Q v+a^{-2}f \right)\,,
\ee
where a scalar function $f$ is related to the perturbation of
$q_i$ in the form
\be
\delta q_i=\partial_i f\,.
\ee
The $\mu=i$ components of Eqs.~(\ref{Tcon1})
and (\ref{Tcon2}) give the Euler equations
\begin{align}
\delta \Lambda&=f+Q\theta\,,
\label{pereq3}\\
\rho_c \dot{\theta}+\rho_c A&=f\,.
\label{pereq4}
\end{align}
Equation (\ref{pereq4}) means that the CDM feels the force $f$
through the interaction with vacuum energy.
{}From Eqs.~(\ref{delLam}) and (\ref{pereq3}), we obtain
\be
f=Q \left( \hat{\theta}-\theta \right)\,,
\ee
which does not vanish for $\hat{\theta} \neq \theta$.
If there is the difference between the four velocities of CDM and
vacuum energy, there is the momentum transfer which affects
the evolution of CDM velocity potential.

We define the CDM density contrast, as
\be
\delta_c \equiv \frac{\delta \rho_c}{\rho_c}\,.
\ee
From Eqs.~(\ref{pereq1})-(\ref{pereq2}) and
(\ref{pereq3})-(\ref{pereq4}), the two variables
$\delta_c$ and $\theta$ satisfy the first-order
differential equations
\begin{align}
\dot{\delta}_c-3\dot{\psi}+\frac{\nabla^2}{a^2}
\left( \theta+a \sigma \right)
&=-\frac{\dot{\delta \Lambda}-\dot{\Lambda}\delta_c}{\rho_c}\,,
\label{peq1} \\
\dot{\theta}+A&=\frac{\delta \Lambda+\dot{\Lambda}
\theta}{\rho_c}\,,
\label{peq2}
\end{align}
where
\be
\sigma \equiv a\dot{E}-B\,.
\ee
The linearly perturbed Einstein equations are given by
$\delta G^{\mu \nu}=8 \pi G (\delta T_m^{\mu \nu}
+\delta T_{\Lambda}^{\mu \nu})$, whose
$(00)$, $(0i)$, and $(ij)$ ($i \neq j$) components lead to
\begin{align}
3H \left( \dot{\psi}+H A \right)-\frac{\nabla^2}{a^2}
\left( \psi+aH \sigma \right)
=&-4\pi G \left( \rho_c \delta_c+\delta \Lambda \right)\,,
\label{peq3}\\
\dot{\psi}+H A=&-4\pi G \rho_c \theta\,,
\label{peq4}\\
a \left( \dot{\sigma}+2H \sigma \right)-A+\psi=&\;0\,.
\label{peq5}
\end{align}
Thus, we derived the full set of linear perturbation equations
(\ref{peq1})-(\ref{peq2}) and (\ref{peq3})-(\ref{peq5})
without choosing particular gauges.
The vacuum fluctuation $\delta \Lambda$, which is
related to its four velocity $\hat{\theta}$ as
Eq.~(\ref{delLam}), can be determined by
specifying a model of interacting vacuum energy \citep{wands2}.
If $\hat{\theta}=\theta$, i.e.,
$\delta \Lambda=-\dot{\Lambda}\theta$,
the right hand-side of Eq.~(\ref{peq2}) vanishes,
with $-\dot{\delta \Lambda}=\ddot{\Lambda}
\theta+\dot{\Lambda}\dot{\theta}$ on
the right hand-side of Eq.~(\ref{peq1}).
In this case, there is no momentum transfer ($\delta q_i=0$)
between CDM and vacuum energy.
This is an example where the perturbation equations
of motion are closed.

We introduce the following gauge-invariant gravitational
potentials \citep{Bardeen},
\be
\Psi=A-\frac{{\rm d}}{{\rm d}t} \left( a \sigma \right)\,,\qquad
\Phi=\psi+aH \sigma\,.
\label{grapo}
\ee
Then, from Eq.~(\ref{peq5}), we obtain
\be
\Psi=\Phi\,,
\label{grare}
\ee
which shows the absence of an anisotropic stress.

Although we have not chosen particular gauges so far,
there are residual gauge degrees of freedom to be fixed.
In the following, we will consider two different gauge choices.

\subsection{Newtonian gauge}

Let us first choose the Newtonian gauge satisfying
\be
B=0\,,\qquad E=0\,,
\ee
under which $\sigma=0$. In this case, the gravitational
potentials (\ref{grapo}) are given by
\be
\Psi=A\,,\qquad \Phi=\psi\,.
\ee
On using the relation (\ref{grare}), Eqs.~(\ref{peq1}), (\ref{peq2}),
(\ref{peq3}), and (\ref{peq4}) reduce, respectively, to
\begin{align}
\dot{\delta}_c-3\dot{\Phi}+\frac{\nabla^2}{a^2} \theta
&=-\frac{\dot{\delta \Lambda}-\dot{\Lambda}\delta_c}{\rho_c}\,,\\
\dot{\theta}+\Phi&=\frac{\delta \Lambda+\dot{\Lambda}
\theta}{\rho_c}\,,\\
3H \left( \dot{\Phi}+H \Phi \right)-\frac{\nabla^2}{a^2} \Phi
&=-4\pi G \left( \rho_c \delta_c+\delta \Lambda \right)\,,\\
\dot{\Phi}+H \Phi&=-4\pi G \rho_c \theta\,.
\end{align}
For a given physical model relating $\delta \Lambda$ with
other perturbations, these equations can be
solved for $\delta_c$, $\theta$, $\Phi$, and $\delta \Lambda$.

\subsection{Synchronous gauge}
\label{synsec}

The synchronous gauge is characterized by the conditions
\be
A=0\,,\qquad B=0\,,
\ee
under which $\sigma=a\dot{E}$.
The gravitational potentials reduce to
\be
\Psi=-\frac{{\rm d}}{{\rm d}t} \left( a^2 \dot{E}
\right)\,,\qquad \Phi=\psi+a^2 H \dot{E}\,.
\ee
We introduce the following combination
\be
h \equiv 2\nabla^2 E-6 \psi\,.
\label{hE}
\ee
Then, Eqs.~(\ref{peq1}), (\ref{peq2}),
(\ref{peq3}), and (\ref{peq4}) yield
\begin{align}
\dot{\delta}_c+\frac{1}{2}\dot{h} +\frac{\nabla^2}{a^2}
\theta &=-\frac{\dot{\delta \Lambda}-\dot{\Lambda} \delta_c}{\rho_c}\,,
\label{pers1}\\
\dot{\theta}&=\frac{\delta \Lambda+\dot{\Lambda}
\theta}{\rho_c}\,,
\label{pers2}\\
\frac{\nabla^2}{a^2} \psi+\frac{1}{2} H \dot{h}
&=4\pi G \left( \rho_c \delta_c+\delta \Lambda \right)\,,
\label{pers3}\\
\dot{\psi}&=-4\pi G \rho_c \theta\,.
\label{pers4}
\end{align}
Equation (\ref{peq5}), which is equivalent to Eq.~(\ref{grare}),
gives the differential equation for $E$, as
\be
a^2 \left( \ddot{E}+3H \dot{E} \right)+\psi=0\,.
\label{Eeq}
\ee
Exerting the operator $\nabla^2$ on Eq.~(\ref{Eeq})
and using Eqs.~(\ref{hE}), (\ref{pers3}), and (\ref{pers4}),
the perturbation $h$ obeys
\be
\ddot{h}+2H\dot{h}+8 \pi G \left( \rho_c \delta_c
-3\rho_c \dot{\theta}+\delta \Lambda
+3\dot{\Lambda} \theta
\right)=0\,.
\label{pers5}
\ee
For a given relation of $\delta \Lambda$ with other perturbations,
we can integrate Eqs.~(\ref{pers1}), (\ref{pers2}), (\ref{pers4}),
(\ref{Eeq}), and (\ref{pers5}) to solve for $\delta_c$, $\theta$, $\psi$,
$E$, and $h$.

\section{Concrete interacting model}
\label{concretesec}

Let us consider a concrete interacting model of vacuum energy and dark matter.
For perturbations, this amounts to giving an explicit relation between
$\delta \Lambda$ and other perturbations.
We study the case in which there is no momentum transfer in
Eq.~(\ref{Qmu2}), i.e.,
\be
q^{\mu}=0\,,
\ee
under which the interaction is restricted to be $Q^{\mu}=Qu^{\mu}$.
In this case we have $\delta q_i=\partial_i f=0$, so
the force $f=Q(\hat{\theta}-\theta)$ exerting on the CDM perturbation
vanishes, i.e., $\hat{\theta}=\theta$.
Then, from Eq.~(\ref{delLam}), the perturbation $\delta \Lambda$
is related to $\theta$, as
\be
\delta \Lambda=-\dot{\Lambda} \theta=Q\theta\,.
\label{nomo}
\ee
Under this condition the right hand side of the Euler Eq.~(\ref{peq2})
is zero, while this is not the case for the terms on the right hand
side of the continuity Eq.~(\ref{peq1}).

Let us derive the perturbation equation of motion for $\delta_c$
in the synchronous gauge.
{}From Eq.~(\ref{pers2}), we obtain $\dot{\theta}=0$, i.e.,
\be
\theta=\theta(x^i)\,,
\label{thecon}
\ee
which means that $\theta$ does not depend on time.
In Fourier space with the comoving wavenumber $k$,
we can write Eq.~(\ref{pers1}) in the form
\be
\dot{\delta}_c+\frac{1}{2}\dot{h}-\frac{k^2}{a^2}\theta
=-\frac{\dot{Q} \theta+Q\delta_c}{\rho_c}\,,
\label{fdelm}
\ee
where we used Eq.~(\ref{nomo}).
We take the time derivative of Eq.~(\ref{fdelm}) and eliminate the
term $2H (k^2/a^2)\theta$ on account of Eq.~(\ref{fdelm}).
On using Eq.~(\ref{pers5}), the CDM density contrast obeys
\begin{align}\label{delmeq}
\ddot{\delta}_c+\left( 2H+\frac{Q}{\rho_c} \right) \dot{\delta}_c
-\left[ 4\pi G \rho_c+\frac{Q^2-(\dot{Q}+5HQ)\rho_c}{\rho_c^2}
\right] \delta_c
+\left[ 8 \pi G Q+\frac{(\ddot{Q}+5H \dot{Q})\rho_c-\dot{Q}Q}
{\rho_c^2} \right] \theta=0\,.
\end{align}
From Eq.~(\ref{fdelm}), the term $(k^2/a^2)\theta$
is at most of the order $\dot{\delta}_c \lesssim H \delta_c$,
and hence $\theta \lesssim a^2 H \delta_c /k^2$.
Moreover, from Eq.~(\ref{cont2}), the coupling $Q$ should be
of order $Q \lesssim 3H \rho_c$.
Then, the ratio between the terms $8\pi G Q \theta$ and
$4\pi G \rho_c \delta_c$ can be estimated as
\be
r \equiv \frac{8\pi G Q \theta}{4\pi G \rho_c \delta_c}
\lesssim \left( \frac{aH}{k} \right)^2\,.
\label{ratio}
\ee
For perturbations deep inside the Hubble
radius ($k \gg aH$), it follows that $r \ll 1$.
Similarly, the other terms appearing as coefficients of
$\theta$ in Eq.~(\ref{delmeq}) are suppressed
relative to the terms proportional to $\delta_c$.
Then, for the modes $k \gg aH$, Eq.~(\ref{delmeq}) is
approximately given by
\begin{align}\label{delmeq2}
\ddot{\delta}_c&+\left( 2H+\frac{Q}{\rho_c} \right) \dot{\delta}_c -\left[ 4\pi G \rho_c+\frac{Q^2-(\dot{Q}+5HQ)\rho_c}{\rho_c^2}
\right] \delta_c \simeq 0\,.
\end{align}
Actually, in view of the time independence of $\theta$ in the synchronous gauge, we can always set it zero, which turns Eq.~(\ref{delmeq2}) an exact result. Note, however, that the sub-horizon approximation is necessary to obtain the same equation in the Newtonian gauge. Although we have chosen the synchronous gauge,
the analysis in the Newtonian gauge gives rise to the
same equation as (\ref{delmeq2}) for the modes
deep inside the Hubble radius.
Introducing the variables
\be
\Gamma_c \equiv \frac{Q}{\rho_c}\,,\qquad
\eta \equiv \int a^{-1}\,{\rm d}t\,,
\label{Gamcdef}
\ee
we can rewrite Eq.~(\ref{delmeq2}) in the form
\begin{align}\label{delmeq3}
\delta_c''&+a (H+\Gamma_c) \delta_c'-\left[ 4\pi G a^2 \rho_c-(a\Gamma_c)'-a^2 \Gamma_c H
\right] \delta_c \simeq 0\,,
\end{align}
where a prime represents the derivative with respect to $\eta$.
The coupling $\Gamma_c$ modifies the growth rate of
$\delta_c$ in comparison to the uncoupled case.

\subsection{Inclusion of baryons}
\label{basec}

So far, we have ignored the perturbation of baryons,
but we would like to take it into account for the J-PAS forecast.
We assume that baryons are coupled to neither vacuum energy
nor dark matter.
Then, the background baryon density
$\rho_b$ obeys the continuity equation
\be
\dot{\rho}_b+3H \rho_b=0\,,
\label{rhob}
\ee
with a vanishing pressure ($p_b=0$).

Let us consider the synchronous gauge with the additional baryon
density contrast $\delta_b$ and velocity field $\theta_b$.
Then, the resulting continuity and Euler equations are
given, respectively, by
\begin{align}
\dot{\delta}_b+\frac{1}{2} \dot{h}+\frac{\nabla^2}{a^2} \theta_b&=0\,,
\label{perb}\\
\dot{\theta}_b&=0\,,
\label{delb}
\end{align}
besides the perturbation Eqs.~(\ref{pers1}) and (\ref{pers2}) for CDM.
The perturbed Einstein Eqs.~(\ref{pers3}) and (\ref{pers4}) are
modified to
\begin{align}
\frac{\nabla^2}{a^2} \psi+\frac{1}{2} H \dot{h}
&=4\pi G \left( \rho_m \delta_m+\delta \Lambda \right)\,,\\
\dot{\psi}&=-4\pi G \left( \rho_c \theta+\rho_b \theta_b \right)\,,
\end{align}
where
\be
\rho_m \delta_m \equiv \rho_c \delta_c+ \rho_b \delta_b\,,
\qquad \rho_m \equiv \rho_c+\rho_b\,,
\label{delm}
\ee
while Eq.~(\ref{Eeq}) is unchanged.
Following the same procedure as that explained in Sec.~\ref{synsec}
and using Eq.~(\ref{delb}), the perturbation $h$, which is
defined in Eq.~(\ref{hE}), obeys
\be
\ddot{h}+2H\dot{h}+8 \pi G \left( \rho_m \delta_m
-3\rho_c \dot{\theta}+\delta \Lambda
+3\dot{\Lambda} \theta
\right)=0\,.
\label{ddoth}
\ee

Now, we focus on the case in which there is no momentum
exchange between vacuum energy and dark matter.
Since the condition (\ref{thecon}) holds, we have
$\dot{\theta}=0$ in Eq.~(\ref{ddoth}).
For perturbations deep inside the Hubble radius,
the terms $\delta \Lambda=Q\theta$ and
$3\dot{\Lambda} \theta =-3Q \theta$ in Eq.~(\ref{ddoth})
are neglected relative to $\rho_c \delta_c$, see Eq.~(\ref{ratio}).
Then, Eq.~(\ref{ddoth}) approximately reduces to
\be
\ddot{h}+2H \dot{h}+8 \pi G \rho_m \delta_m
\simeq 0\,.
\ee
Differentiating Eqs.~(\ref{pers1}) and (\ref{perb}) with
respect to $t$ and taking the similar procedure to that
explained in Sec.~\ref{synsec},
the CDM and baryon density contrasts in Fourier space
obey the following differential equations
\begin{align}
\ddot{\delta}_c&+\left( 2H+\frac{Q}{\rho_c} \right) \dot{\delta}_c -4\pi G \rho_m \delta_m -\frac{Q^2-(\dot{Q}+5HQ)\rho_c}{\rho_c^2}\delta_c
\simeq 0\,,
\label{delmf} \\
\ddot{\delta}_b&+2H \dot{\delta}_b-4\pi G \rho_m \delta_m
\simeq 0\,,
\end{align}
which are valid for the modes $k \gg aH$.
By using the variables $\Gamma_c$ and $\eta$ defined
in Eq.~(\ref{Gamcdef}), these equations can be expressed as
\begin{align}\label{delmfi}
\delta_c''&+a (H+\Gamma_c) \delta_c'
-4\pi G a^2 \rho_m \delta_m +\left[ (a\Gamma_c)'+a^2 \Gamma_c H
\right] \delta_c  \simeq 0\,,
\end{align}
\begin{align}\label{delbfi}
\delta_b''+a H \delta_b'-4\pi G a^2 \rho_m
\delta_m \simeq 0\,.
\end{align}
Since $|\rho_c \delta_c| \gg |\rho_b \delta_b|$, the baryon
density contrast grows under the influence of gravitational potentials sourced
mostly by the CDM density perturbation. There is no scale-dependence in Eqs.~(\ref{delmfi}) and (\ref{delbfi}) governing the evolution of perturbations on sub-horizon scales, whose property is attributed to the absence of pressures for CDM and baryons. In our interacting theory, the dark energy pressure does not affect the sound speeds of CDM and baryons either.


We can also derive the perturbation equations of motion for the total density matter contrast $\delta_m$ defined in Eq.~(\ref{delm}).
On using the approximation similar to Eq.~(\ref{ratio})
for the modes deep inside the Hubble radius, we obtain
\begin{align}
\ddot{\delta}_m+\left( 2H+\frac{Q}{\rho_m} \right) \dot{\delta}_m
-\left[ 4\pi G \rho_m+\frac{Q^2-(\dot{Q}+5HQ)\rho_m}{\rho_m^2}
\right] \delta_m
\simeq Q \frac{k^2}{a^2} \frac{\rho_b}{\rho_m^2}
\left( \theta-\theta_b \right)\,.
\end{align}
In terms of the conformal time $\eta$, this can be expressed as
\begin{align}\label{delmg}
\delta_m''+a \left( H+\Gamma \right)\delta_m'
-\left[ 4\pi G a^2 \rho_m- \left( a \Gamma \right)'
-a^2 \Gamma H \right] \delta_m
\simeq Q k^2 \frac{\rho_b}{\rho_m^2}
\left( \theta-\theta_b \right)\,,
\end{align}
where
\be
\Gamma \equiv \frac{Q}{\rho_m}\,.
\label{Gamdef}
\ee
The right hand side of Eq.~(\ref{delmg}) does not vanish for
$\theta \neq \theta_b$. This means that the momentum exchange
between CDM and baryons can affect the dynamics of $\delta_m$.
Since both $\theta$ and $\theta_b$ are constants in time,
this term vanishes for the initial condition $\theta=\theta_b$.
In this case, Eq~(\ref{delmg}) reduces to the second-order
differential equation of $\delta_m$.

\subsection{Background}
\label{backsec}

Let us consider a possible choice of the interacting function $Q$
to study the background cosmological evolution.
We take into account the baryons whose background density $\rho_b$
obeys the continuity Eq.~(\ref{rhob}).
The densities of vacuum energy and CDM satisfy Eqs.~(\ref{cont1})
and (\ref{cont2}). In the presence of baryons, Eqs.~(\ref{back1})
and (\ref{back2}) are modified to
\begin{align}
3H^2 &= 8\pi G \left( \rho_m+\Lambda \right)\,,
\label{back1b}\\
\dot{H} &= -4\pi G \rho_m \,,
\label{back2b}
\end{align}
where $\rho_m=\rho_c+\rho_b$.
For the interacting vacuum energy, we choose a function of the form \citep{Pigozzo}
\be
\Lambda=\frac{\sigma H^{-2\alpha}}{8\pi G}\,,
\label{asu}
\ee
where $\sigma~(>0)$ and $\alpha~(>-1)$ are constants.
{}From Eq.~(\ref{back1b}), there is the relation
\be
\sigma=3 \left( 1-\Omega_m \right)
H_0^{2(\alpha+1)}\,,
\label{sigma}
\ee
where $\Omega_m=8\pi G \rho_{m0}/(3H_0^2)$
is today's density parameter of total non-relativistic matter
(the subscript ``0'' represents today's values).
On using Eqs.~(\ref{back1b}), (\ref{back2b}), and (\ref{asu}),
the Hubble parameter obeys the differential equation
\be
\frac{{\rm d}H}{{\rm d}z}=
\frac{3H}{2(1+z)} \left[ 1 - \left( 1-\Omega_m \right)
\left( \frac{H}{H_0} \right)^{-2(\alpha+1)} \right]\,,
\label{dH}
\ee
where $z=1/a-1$ is the redshift, and we used the
relation ${\rm d}z/{\rm d}t=-(1+z)H$.
Integrating Eq.~(\ref{dH}) with respect to $z$, we obtain
\be
H(z)=H_0 \left[ 1-\Omega_m +\Omega_m
\left( 1+z \right)^{3(\alpha+1)}
\right]^{1/[2(\alpha+1)]}\,,
\label{H}
\ee
which satisfies $H=H_0$ at $z=0$.
The decay rate (\ref{Gamdef}) is given by
\be
\Gamma=-\frac{\dot{\Lambda}}{\rho_m}
=\frac{\alpha \sigma H^{-2\alpha-1}}{4\pi G}
\frac{\dot{H}}{\rho_m}
=-\alpha \sigma H^{-2\alpha-1} \,.
\ee

The non-interacting cosmological constant
corresponds to $\alpha = 0$, in which case $\Gamma=0$.
For $\alpha<0$, the creation of CDM occurs through the positive
coupling $\Gamma$.
In particular, $\Gamma$ is constant for $\alpha = -1/2$.
From (\ref{H}) we observe that the ansatz (\ref{asu}) corresponds to a decomposed, non-adiabatic generalised Chaplygin gas \citep{wands21,wands22,wands23,non-adiabatic1,non-adiabatic2,wang,Marttens:2017njo,alemaes},
which behaves like conserved matter at high redshifts and approaches
a cosmological constant in the asymptotic future.
The late-time non-adiabaticity prevents oscillations and instabilities in the power spectrum, contrary to what happens in adiabatic versions of the generalised Chaplygin gas \citep{gCg1,gCg2,gCg3,gCg4,gCg5,gCg6,gCg7,gCg8,oliver}.
This parametrization does not encompass all possible forms of interactions, but it can be used to search for signatures of
the interacting vacuum energy in current and future observations.

The baryon density has the following redshift dependence,
\be
\rho_b(z)=\frac{3H_0^2}{8\pi G} \Omega_b (1+z)^3\,,
\label{rhobz}
\ee
where $\Omega_b=8 \pi G\rho_{b0}/(3H_0^2)$.
On using Eqs.~(\ref{back2b}), (\ref{H}), and
Eq.~(\ref{rhobz}), the CDM density $\rho_c=\rho_m-\rho_b$
is given by the form
\begin{align}\label{rhocz}
\rho_c (z)=\frac{3H_0^2}{8\pi G} \left( 1+z \right)^3
\left[ \left( 1+z \right)^{3\alpha} \Omega_m
\left\{ 1-\Omega_m +\Omega_m \left( 1+z \right)^{3(\alpha+1)}
\right\}^{-\alpha/(\alpha+1)} -\Omega_b \right]\,.
\end{align}
At high redshifts ($z \gg 1$), it follows that
\be
\rho_c (z) \simeq \frac{3H_0^2}{8\pi G} \left( 1+z \right)^3
\left[ \Omega_m^{1/(\alpha+1)}-\Omega_b \right]\,.
\ee
To avoid a negative CDM density, we require that $\rho_c (z)>0$.
This condition translates to
\be
\alpha> \alpha_c \equiv -1+\frac{\ln \Omega_m}{\ln \Omega_b}\,.
\label{conal}
\ee
When $\Omega_m=0.32$ and $\Omega_b=0.05$, for example,
we have $\alpha_c=-0.62$.
At low redshifts around $z=0$, the expansion of
Eq.~(\ref{rhocz}) shows that, as long as $\Omega_m>\Omega_b$,
the leading-order term of $\rho_c(z)$ is positive.
For theoretical consistency, $\alpha$ should be in the
range (\ref{conal}).

\subsection{Growth rate of perturbations}
\label{growsec}

\begin{figure}
\begin{center}
\includegraphics[height=3.2in,width=3.2in]{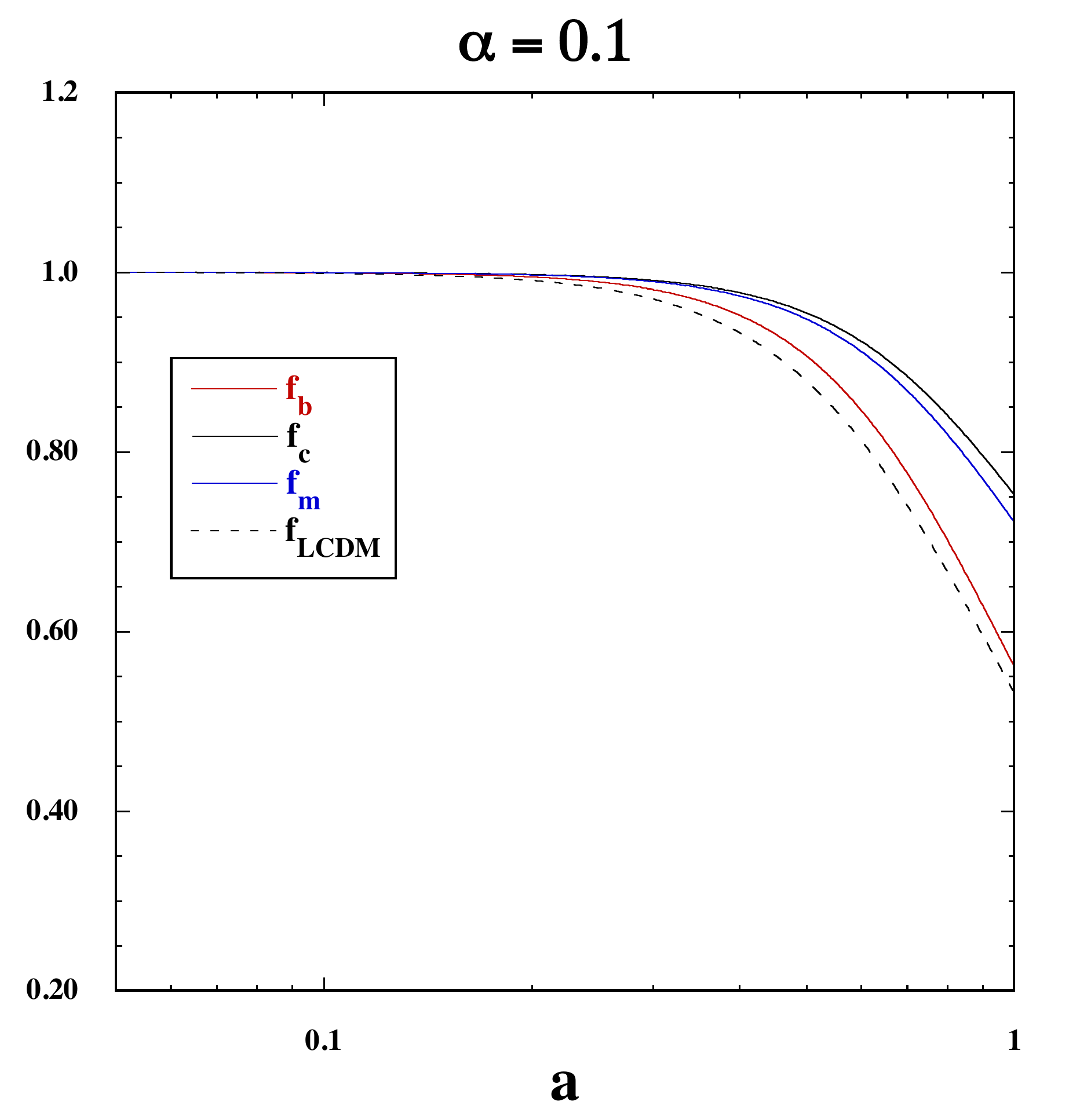}
\includegraphics[height=3.2in,width=3.2in]{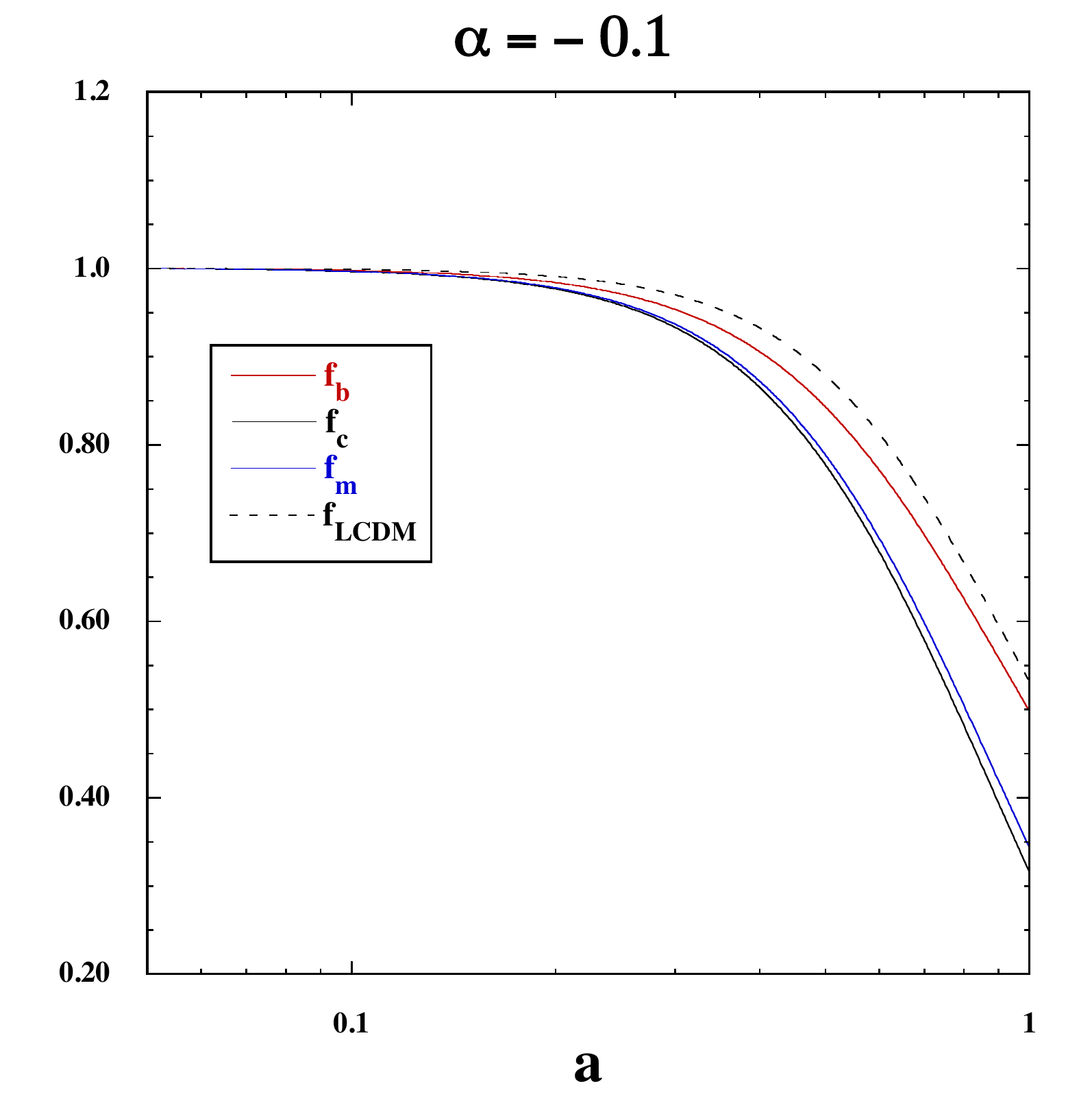}
\end{center}
\caption{\label{ffig}Evolution of $f_b$, $f_c$, and $f_m$ versus the scale factor $a$
for $\alpha=0.1$ (left) and $\alpha=-0.1$ (right), respectively.
We also plot the evolution of $f_b~(=f_c=f_m)$ in the $\Lambda$CDM
model as a dashed curve. Today's values of the density parameters
are chosen to be $\Omega_b=0.05$ and
$\Omega_m=0.32$.}
\end{figure}

We define the functions describing the growth rates of CDM,
baryons, and total matter, as
\be
f_i(a) \equiv \frac{{\rm d} \ln \delta_{i}(a)}{{\rm d} \ln a}
=\frac{\dot{\delta}_i}{H \delta_{i}}\,,
\ee
where $i=b, c, m$, respectively.
For sub-horizon perturbations, these functions are known
by integrating Eqs.~(\ref{delmfi}) and (\ref{delbfi})
together with the relation (\ref{delm}).
In Fig.~\ref{ffig}, we plot the evolution of $f_b$, $f_c$, and $f_m$
for $\alpha=0.1$ (left) and $\alpha=-0.1$ (right).
We choose the initial conditions of density contrasts same as
those in the $\Lambda$CDM model ($\alpha=0$), i.e.,
$f_b=1$ and $f_c=1$ at $a=0.05$.
For the purpose of showing the evolution of $f_i$,
it is sufficient to choose an arbitrary amplitude
satisfying $\delta_b=\delta_c$.

In Fig.~\ref{ffig}, the evolution of growth rates
in the $\Lambda$CDM model is also shown, in which case
$f_b=f_c=f_m$ due to the absence of interactions and
the choice of same initial conditions for $\delta_b$ and $\delta_c$.
At late times the interacting vacuum energy
model with $\alpha>0$ leads to the value of $f_c$ larger than
in the $\alpha=0$ case, while for $\alpha<0$,
$f_c$ is subject to suppression.
This behavior can be understood by expressing
Eq.~(\ref{delmfi}) in the form
\be
\delta_c''+a (H+\Gamma_c) \delta_c'
-4\pi a^2 \left(  G_c \rho_c \delta_c
+G \rho_b \delta_b \right)
\simeq 0\,,
\label{delmfi2}
\ee
where $G_c$ is the effective gravitational coupling for $\delta_c$
defined by
\be
G_c \equiv G-\frac{\dot{\Gamma}_c+2H \Gamma_c}{4\pi \rho_c}\,.
\ee
Assuming that $|\alpha| \ll 1$ and expanding $G_c$ around $\alpha=0$,
we obtain
\be
G_c=G \left[ 1+\alpha (1-\Omega_m) \frac{\tilde{\Omega}_m
(3\tilde{\Omega}_m+4)}{\tilde{\Omega}_c^2}
\frac{H_0^2}{H^2}+{\cal O} (\alpha^2) \right]\,,
\ee
where $\tilde{\Omega}_m=8\pi G \rho_m/(3H^2)$
and $\tilde{\Omega}_c=8\pi G \rho_c/(3H^2)$.
For $\alpha>0$, we have $G_{c}>G$ and hence the growth of $\delta_c$
is enhanced in comparison to the $\Lambda$CDM model.
For $\Omega_m=0.32$ and $\Omega_c=0.27$, today's value of
$G_c$ is given by $G_c \simeq G(1+14.8 \alpha)$.
This means that, even for $|\alpha|={\cal O}(0.1)$,
$G_c$ is significantly modified relative to the $\alpha=0$ case.
For $\alpha>0$, the term $a\Gamma_c$ in front of
$\delta_c'$ in Eq.~(\ref{delmfi2}) is negative, so this also
works to enhance the growth rate of $\delta_c$.
The enhancement of $\delta_c$ also leads to the larger baryon
growth rate $f_b$ through Eq.~(\ref{delbfi}) in comparison
to the $\alpha=0$ case.
As we see in the left panel of Fig.~\ref{ffig},
$f_c$ is larger than $f_b$
(see Ref.~\citep{velten} for a related work).
The evolution of total matter growth
function is similar to that of CDM, but $f_m$
is slightly smaller than $f_c$ due to the presence of baryons.

For $\alpha<0$, the growth of CDM density contrast is suppressed
in comparison to the $\alpha=0$ case. As we observe in
the right panel of Fig.~\ref{ffig}, the suppression of $f_c$
also gives rise to the values of $f_b$ and $f_m$ smaller than those
in the $\Lambda$CDM model. Thus, the large or small cosmic growth
rates can be realized in the interacting vacuum energy scenario, depending
on the sign of $\alpha$. This allows us to probe the signature of
interactions observationally.

In RSD measurements, the matter velocity potential is used to quantify the
galaxy distortion in redshift space.
In Eq.~(\ref{fdelm}) the term $-\dot{Q} \theta/\rho_c$ is neglected
relative to the other terms for sub-horizon perturbations, so using the CDM
growth rate $f_c$ leads to
\be
H \left( f_c+g_c \right) \delta_c =\frac{k^2}{a^2}\theta-\frac{1}{2} \dot{h}\,,
\ee
where $g_c \equiv Q/(H \rho_c)$.
This means that the CDM growth rate associated with
the velocity potential $\theta$
corresponds to \citep{borges-wands,borges-carneiro}
\be
f_{c,{\rm RSD}}=f_c+g_c\,.
\label{fgc}
\ee
Hence there is the additional contribution to $f_c$ from
the coupling $Q$.
{}From Eq.~(\ref{perb}), the growth rate of baryon density contrast
relevant to RSD measurements is simply given by $f_b$.
As our tracers are luminous baryonic matter, we mostly use $\delta_b$ in
our forecast analysis.
However, the late-time enhancement or suppression
in $\delta_m$ can lead to tight observational
constraints on the interaction between vacuum energy and CDM,
so we will also discuss the case of total matter
density contrast.

\section{Matter power spectrum}
\label{Fishersec}

To confront the interacting vacuum energy model with the observations
of luminous galaxies, we define the matter power spectrum
in Fourier space at a redshift $z$, as
\begin{equation}
\mathcal{P}_{L}(k,z) = \mathcal{P}_{L,0}(k)
\left(\frac{\delta_{b,m}(a)}{\delta_{b,m}(1)}\right)^2\,,
\end{equation}
where we exploited the fact that Eq.~(\ref{delbfi}) does not
contain the $k$ dependence, and the suffices $b$ and $m$ refer to the baryonic and total matter scenarios which we consider separately in this work.
The scale-dependent part $\mathcal{P}_{L,0}(k)$ is
today's matter power spectrum, which is given by
\begin{equation}
\mathcal{P}_{L,0}(k)=\mathcal{P}_{0} k^{n_s} \mathcal{T}^{2}(k)\,,
\label{PL0}
\end{equation}
where the scaling constant $\mathcal{P}_{0}$ is found using
the usual normalization of $\sigma_8$, i.e.,
\begin{equation}
\sigma^{2}_{8,0} = \frac{1}{2 \pi^{2}} \int^{\infty}_{0}
k^{2} W^{2}(k,R) \mathcal{P}_{L,0}(k)\, {\rm d} k\,,
\end{equation}
with the window function
\begin{equation}
W(k,R) = \frac{3}{k^3\,R^3} \left[ \sin(k R) -k R \cos(k R) \right]\,,
\end{equation}
at the comoving scale $R = 8\,h^{-1}$ Mpc.
In Eq.~(\ref{PL0}), the scale dependence is present in the
spectral index $n_s$ of primordial scalar perturbations and the transfer
function $\mathcal{T}$. The latter accommodates the evolution
of gravitational potentials from the radiation dominance
to the matter era.

The galaxy linear power spectrum is defined
as\footnote{As we showed in Eq.~(\ref{fgc}), the CDM power spectrum
can be derived by performing the concomitant substitution
$f_b \rightarrow f_c + g_c$ in Eq.~(\ref{eq:power_v0}).}
\begin{equation}
\label{eq:power_v0}
\mathcal{P}_{g}(k,z,\mu) = \left[ b_{s}(z) + f_{s}(z)\mu^{2} \right]^{2}
\frac{\mathcal{P}_{L,0}(k)}{\sigma^{2}_{8,0}}
\exp \left[ - k^{2} \mu^{2}\Sigma^{2}_{z}(z) \right]\,,
\end{equation}
where $b_{s}(z) = b_{g}(z) \sigma_{8}(z)$ and
$f_{s}(z) = f_b(z) \sigma_{8}(z)$, with $b_g$ and $f_b$ being
the galaxy bias and the baryonic growth rate, respectively.
The cosine of the angle of unit wavevector
${\bm k}$ with respect to the line-of-sight direction is written as
$\mu={\bm k} \cdot {\bm r}/r$.
In Eq.~(\ref{eq:power_v0}), we have inserted the damping factor
$\exp \left[ - k^{2} \mu^{2}\Sigma^{2}_{z}(z) \right]$ to account
for redshift uncertainties $\sigma_{z}(z)$,
weighed by $\Sigma_{z}(z)=\sigma_{z}(z)/H(z)$.
Finally, the observed matter power spectrum can be written as
\begin{equation}
\mathcal{P}_{\rm obs}(k,z,\mu)=
\mathcal{P}_{g}(k,z,\mu) + \mathcal{P}_{\rm shot}(z)\,,
\end{equation}
where the shot noise is defined by $\mathcal{P}_{\rm shot}(z) = 1/n(z)$,
with $n$ being the comoving galaxy density per redshift bin.

For the damping of non-linear evolution of the matter power spectrum,
we follow the approach given in Refs.~\citep{Eisenstein:2006nj,Seo:2007ns,Font-Ribera:2013rwa}.
On scales larger than $100\,h^{-1}$ Mpc, i.e., $k \lesssim 0.06\,h$ Mpc$^{-1}$,
which is our range of interest, the non-linear evolution leads to a damping/suppression
of all linear theory information, as shown in the left panel of Fig.~1 of Ref.~\citep{Crocce:2007dt}
and discussed extensively in Refs.~\citep{Eisenstein:2006nj,Seo:2007ns,Crocce:2007dt,White:2010qd,Font-Ribera:2013rwa,Amendola:2016saw}.
Such damping can be modelled by the non-linear
power spectrum \citep{Eisenstein:2006nj,Seo:2007ns,Font-Ribera:2013rwa,Amendola:2016saw}
\begin{align}
\mathcal{P}_{\rm NL}(k,z,\mu)&=\mathcal{P}_{\rm obs}(k,z,\mu)
\exp \left[ -\frac{k^{2}}{2} (1-\mu^{2})\Sigma^{2}_{\perp}(z) -\frac{k^{2}}{2} \mu^{2}\Sigma^{2}_{||}(z) \right]\;,
\end{align}
where the damping factors are needed to take into account the smearing
due to non-linear structure formation along $(\Sigma_{||})$
and across $(\Sigma_{\perp})$ the line of
sight \citep{Eisenstein:2006nj,Seo:2007ns}.
The damping factors are given by
\begin{align}
\Sigma_{\perp}(z) &= 0.785\, \Sigma_{0}\, \frac{\delta_{b,m}(z)}{\delta_{b,m}(0)}\,,\\
\Sigma_{||}(z) &= \left[1+f_b(z)\right] \Sigma_{\perp}(z)\,,
\end{align}
where $\Sigma_{0} = 11 h^{-1} \, \text{Mpc}$. One should point out that the numerical factors which appear in the previous expressions are technically model dependent, i.e. for the interacting vacuum energy models they might be different from the case of a standard $\Lambda$CDM. Although, it is generally assumed in the literature that their impact on the final Fisher analysis is mild, for they do not enter in the derivatives of the power spectrum with respect to the model parameters, thus effectively resulting in a sort of marginalization procedure \citep{Seo:2007ns}.

\section{Forecasts}
\label{resultsec}

We exploit a Fisher Matrix to compute errors in the J-PAS forecast.
In the case of a single tracer, a generic Fisher Matrix element, $F_{ij}$,
can be calculated as \citep{amendola2010dark}
\begin{align}
\label{eq:Fisher}
F_{ij}(z) &= \frac{1}{4 \pi^{2}} \int^{1}_{-1} {\rm d}\mu' \int^{k_{\rm max}}_{k_{\rm min}}
{\rm d}k' k'^{2}\frac{V_{\rm eff}(k',z,\mu')}{2} \frac{{\rm d}
\ln \mathcal{P}_{\rm obs}(k',z,\mu')}{{\rm d} p_{i}}
\frac{{\rm d} \ln \mathcal{P}_{\rm obs}(k',z,\mu')}{{\rm d} p_{j}} \\
&\times \exp \left[ -k'^{2} (1-\mu'^{2})\Sigma^{2}_{\perp}(z) -k'^{2}
\mu'^{2}\Sigma^{2}_{||}(z) \right]\,, \nonumber
\end{align}
where
$k_{\rm min} = 0.007\,h \, \mathrm{Mpc}^{-1}$ and
$k_{\rm max} \to \infty$~\citep{Amendola:2013qna}.
The effective volume of the survey is
\begin{equation}
V_{\rm eff}(k,z,\mu) = V_{a}(z)
\left[ \frac{n(z) \mathcal{P}_{g}(k,z,\mu)}{n(z) \mathcal{P}_{g}(k,z,\mu) + 1}
\right]^{2}\, ,
\end{equation}
where
\begin{equation}
V_{a}(z) = \frac{4\pi f_{\rm sky}}{3}
\left[ \chi(\bar{z}_a)^3 - \chi(\bar{z}_{a-1})^3 \right]
\end{equation}
is the volume span by the survey in each redshift bin, and
\begin{equation}
\chi(z) = \int_{0}^{z}{\frac{{\rm d}z'}{H(z')}}\;
\label{chi}
\end{equation}
is the comoving distance with $\bar{z}_a$ being
the upper limit of $a$-th redshift bin.

{\renewcommand{\tabcolsep}{2.0mm}
{\renewcommand{\arraystretch}{1.5}
\begin{table*}
\centering
\begin{minipage}{0.7\textwidth}
\caption{Surveys specifications.
Column $1$: name of the survey;
column $2$: redshift range;
column $3$: redshift bin width;
column $4$: survey area in sq. deg.;
column $5$: redshift error;
column $6$: galaxy bias functions \protect\footnote{Note that the provided galaxy bias functions should be considered model dependent, thus for our interacting vacuum energy model they might be different. But the impact on the final Fisher analysis is smeared out by performing a marginalization over these functions, as described in the following pages.}.}\label{tab:surveys}
\resizebox*{\textwidth}{!}{
\begin{tabular}{cccccc}
\hline
\hline
 & $(z_{\rm min},z_{\rm max})$ & $\Delta z$ & $A_{{\rm survey}}$  & $\sigma_{z}$  & $b_{g}$  \\
 &                     &            & (sq. deg.)   &               &          \\
\hline
DESI-BCG & $(0.05,0.6)$ &  \multirow{4}{*}{$0.2$} & \multirow{3}{*}{$14000$} & \multirow{3}{*}{$0.0005\ (1+z)$}& $1.34\frac{\delta_{m,b}(0)}{\delta_{m,b}(z)}$ \\
DESI-LRG & $(0.6,1.2)$  &                         &                          &                                 & $1.7\frac{\delta_{m,b}(0)}{\delta_{m,b}(z)}$ \\
DESI-ELG & $(0.6,1.8)$  &                         &                          &                                 & $0.84\frac{\delta_{m,b}(0)}{\delta_{m,b}(z)}$ \\
DESI-QSO & $(0.6,1.8)$  &                         &                          & $0.001\ (1+z)$                  & $0.53+0.289 (1+z)^2$ \\
\hline
\hline
\textit{Euclid}     & $(0.9,1.8)$ & $0.2$ & $15000$ & $0.001\ (1+z)$ & Table 3 in \citep{EucFish} \\
\hline
J-PAS-LRG & $(0.2,1.2)$ & \multirow{3}{*}{$0.2$} & \multirow{3}{*}{$4000$ - $8500$} & \multirow{3}{*}{$0.003\ (1+z)$} & $1.7\frac{\delta_{m,b}(0)}{\delta_{m,b}(z)}$ \\
J-PAS-ELG & $(0.2,1.4)$ &                        &                                  &                              & $0.84\frac{\delta_{m,b}(0)}{\delta_{m,b}(z)}$ \\
J-PAS-QSO & $(0.2,4)$ &                        &                                  &                              &  $0.53+0.289 (1+z)^2$ \\
\hline
\hline
\end{tabular}}
\end{minipage}
\end{table*}}}

{\renewcommand{\tabcolsep}{2.0mm}
{\renewcommand{\arraystretch}{1.5}
\begin{table*}
\centering
\begin{minipage}{0.8\textwidth}
\caption{Fisher Matrix scenarios.
Column $1$: step in Fisher Matrix calculation;
column $2$: name of Fisher Matrix;
column $3$: parameters of the Fisher Matrix; column $4$: comments.}\label{tab:fisher_scenarios}
\resizebox*{\textwidth}{!}{
\begin{tabular}{cccc}
\hline
\hline
\multirow{3}{*}{0} & \multirow{3}{*}{$\mathcal{F}^{\rm bin}$} & $\{\ln D_A, \ln H, f_s, b_s,
\mathcal{P}_{{\rm shot}}, \Omega_m, h, n_s \}$ & single tracer: $8 \times 8$ matrix\\
                   &               & $\{\ln D_A, \ln H, f_s, b_{s_{1}}, b_{s_{2}}, \mathcal{P}_{\rm shot},
                   \Omega_m, h, n_s \}$ & $2$ tracers: $9 \times 9$ matrix \\
                   &               & $\{\ln D_A, \ln H, f_s, b_{s_{1}}, b_{s_{2}}, b_{s_{3}},
                  \mathcal{P}_{\rm shot}, \Omega_m, h, n_s \}$ & $3$ tracers: $10 \times 10$ matrix \\
\hline
\hline
\multicolumn{4}{c}{BAO} \\
\multirow{2}{*}{A1} & \multirow{2}{*}{$\mathcal{F}^{\rm bin}_{\rm marg}$} & \multirow{2}{*}{$\{\ln D_A, \ln H \}$}
& marginalization over $\{f_s, b_{s_{i}}, \mathcal{P}_{\rm shot}, \Omega_m, h, n_s \}$\\
                    &              &  & $2 \times 2$ matrix \\
\hline
\multirow{3}{*}{A2} & \multirow{3}{*}{$\mathcal{F}^{\rm bin}_{\rm cosmo}$}
& \multirow{3}{*}{$\{\Omega_m, \alpha \}$} & projection of $\mathcal{F}^{\rm bin}_{\rm marg}$
onto cosmological parameters \\
                    &              &  & $2 \times 2$ matrix \\
                    &              &  & redshift rows of Tables~\ref{tab:results_EUCLID-BAO}~-~\ref{tab:results_DESI-BAO}~-~\ref{tab:results_JPAS4000-BAO}~-~\ref{tab:results_JPAS8500-BAO} \\
\hline
\multirow{3}{*}{A3} & \multirow{3}{*}{$\mathcal{F}^{\rm tot}_{\rm cosmo}$} &
\multirow{3}{*}{$\{\Omega_m, \alpha \}$} & sum of $\mathcal{F}^{\rm bin}_{\rm cosmo}$  \\
                    &              &  & $2 \times 2$ matrix \\
                    &              &  & last row of Tables~\ref{tab:results_EUCLID-BAO}~-~
\ref{tab:results_DESI-BAO}~-~\ref{tab:results_JPAS4000-BAO}~-~\ref{tab:results_JPAS8500-BAO} \\
\hline
\hline
\multicolumn{4}{c}{BAO+RSD} \\
\multirow{2}{*}{B1} & \multirow{2}{*}{$\mathcal{F}^{\rm bin}_{\rm marg}$} &
\multirow{2}{*}{$\{\ln D_A, \ln H, f_s \}$} & marginalization over $\{b_{s_{i}}, \mathcal{P}_{\rm shot},
\Omega_m, h, n_s \}$\\ &          &  & $3 \times 3$ matrix \\
\hline
\multirow{3}{*}{B2} & \multirow{3}{*}{$\mathcal{F}^{\rm bin}_{\rm cosmo}$} &
\multirow{3}{*}{$\{\Omega_m, \alpha \}$} & projection of $\mathcal{F}^{\rm bin}_{\rm marg}$
onto cosmological parameters \\
                    &              &  & $2 \times 2$ matrix \\
                    &              &  & redshift rows of Tables~\ref{tab:results_EUCLID-BAO+RSD}~-~\ref{tab:results_DESI-BAO+RSD}~-~\ref{tab:results_JPAS4000-BAO+RSD}~-~\ref{tab:results_JPAS8500-BAO+RSD} \\
\hline
\multirow{3}{*}{B3} & \multirow{3}{*}{$\mathcal{F}^{\rm tot}_{\rm cosmo}$} & \multirow{3}{*}{$\{\Omega_m, \alpha \}$} & sum of $\mathcal{F}^{\rm bin}_{\rm cosmo}$  \\
                    &              &  & $2 \times 2$ matrix \\
                    &              &  & last row of
                                   Tables~\ref{tab:results_EUCLID-BAO+RSD}~-~\ref{tab:results_DESI-BAO+RSD}~-~\ref{tab:results_JPAS4000-BAO+RSD}~-~\ref{tab:results_JPAS8500-BAO+RSD}
                                   \\
\hline
\hline
\multicolumn{4}{c}{BAO+RSD+PS} \\
\multirow{2}{*}{C1} & \multirow{2}{*}{$\mathcal{F}^{\rm bin}_{\rm marg}$} &
\multirow{2}{*}{$\{\ln D_A, \ln H, f_s,\Omega_m, h, n_s \}$} & marginalization over $\{b_{s_{i}}, \mathcal{P}_{\rm shot}\}$\\
                    &              &  & $6 \times 6$ matrix \\
\hline
\multirow{3}{*}{C2} & \multirow{3}{*}{$\mathcal{F}^{\rm bin}_{\rm cosmo}$} &
\multirow{3}{*}{$\{\Omega_m, \alpha \}$} & projection of $\mathcal{F}^{\rm bin}_{\rm marg}$ onto cosmological parameters \\
                    &              &  & $2 \times 2$ matrix \\
                    &              &  & redshift rows of Tables~\ref{tab:results_EUCLID-FULL}~-~\ref{tab:results_DESI-FULL}~-~\ref{tab:results_JPAS4000-FULL}~-~\ref{tab:results_JPAS8500-FULL} \\
\hline
\multirow{3}{*}{C3} & \multirow{3}{*}{$\mathcal{F}^{\rm tot}_{\rm cosmo}$} & \multirow{3}{*}{$\{\Omega_m, \alpha \}$} &
sum of $\mathcal{F}^{\rm bin}_{\rm cosmo}$  \\
                    &              &  & $2 \times 2$ matrix \\
                    &              &  & last row of
Tables~\ref{tab:results_EUCLID-FULL}~-~\ref{tab:results_DESI-FULL}
~-~\ref{tab:results_JPAS4000-FULL}~-~\ref{tab:results_JPAS8500-FULL}
                                   \\
\hline
\hline
\end{tabular}}
\end{minipage}
\end{table*}}}

Since the surveys considered in our analysis will have more than one tracer with different biases used to probe the same patch of the sky at the same redshift range, we also take into account the cross correlation and define a generalized
Fisher Matrix as \citep{Abramo:2013awa,Zhao:2015gua}
\begin{align}
\label{eq:Fisher-multi}
F_{ij}(z) &= \sum^{N}_{X,Y=1}\frac{1}{4 \pi^{2}}
\int^{1}_{-1} {\rm d}\mu' \int^{k_{\rm max}}_{k_{\rm min}} {\rm d}k' k'^{2}
V_{a}(z) \frac{{\rm d} \ln \widehat{P}_{X,{\rm obs}}(k',z,\mu')}
{{\rm d} p_{i}} \widehat{F}_{XY}\frac{{\rm d} \ln \widehat{P}_{Y,{\rm obs}}(k',z,\mu')}
{{\rm d} p_{j}} \\
&\times \exp \left[ -k'^{2} (1-\mu'^{2})\Sigma^{2}_{\perp}(z)
-k'^{2} \mu'^{2}\Sigma^{2}_{||}(z) \right]\,, \nonumber
\end{align}
where the indices $(X,Y)$ run over the galactic tracers, i.e., LRG, ELG,
and QSO, and $N=2$ or $3$ depending on
the tracer (see Table \ref{tab:surveys} for survey specifications).
Also, the dimensionless effective power
$\widehat{P}_{X,{\rm obs}}$ is defined as
\begin{equation}
\widehat{P}_{X,{\rm obs}}
= n_{X} \mathcal{P}_{X,{\rm obs}}\, ,
\end{equation}
where $n_{X}$ is the comoving galaxy density per redshift bin
per galaxy population and the Fisher information density $\widehat{F}_{XY}$ is written as
\begin{equation}
\widehat{F}_{XY} = \frac{1}{4} \left[ \delta_{XY} \frac{\widehat{P}_{X,{\rm obs}}
\widehat{P}_{{\rm obs}}}{1+\widehat{P}_{{\rm obs}}}
+ \frac{\widehat{P}_{X,{\rm obs}}\widehat{P}_{Y,{\rm obs}}
(1-\widehat{P}_{{\rm obs}})}{(1+\widehat{P}_{{\rm obs}})^{2}}\right]
\end{equation}
with $\widehat{P}_{{\rm obs}} = \sum_{X} \widehat{P}_{X,{\rm obs}}$.
Clearly, for the single-tracer case, Eq.~(\ref{eq:Fisher-multi})
reduces to Eq.~(\ref{eq:Fisher}).

Adopting the nomenclature of Ref.~\citep{Wang:2010gq},
we consider the following combinations in our analysis (also summarised in Table~\ref{tab:fisher_scenarios}):
\begin{itemize}
 \item ``$\mathcal{P}(k)$-marginalised-over-shape'' Fisher Matrix, where only BAO information is taken into account (henceforth ``BAO'');
 \item ``$\mathcal{P}(k)$-marginalised-over-shape'' Fisher Matrix, with both BAO and RSD included
(henceforth ``BAO+RSD'');
 \item ``full ${\cal P}(k)$ method with growth information included'' Fisher Matrix, where the power spectrum broadband ``shape-parameters'' are added to BAO and RSD (henceforth ``BAO+RSD+PS'').
\end{itemize}
In order to avoid numerical instabilities, we follow the procedure reported in Ref.~\citep{JDEM}
for marginalisation\footnote{Note that the shot noise $P_{\rm shot}$ makes the matrix ill-conditioned.
After applying the check as in Eq.~(14) of Ref.~\citep{JDEM}, we have effectively verified
that the corresponding eigenvalue is much smaller than those related to the other parameters,
and that the corresponding row is near zero. Thus, the shot noise component is uncorrelated with
the parameters we are interested in, and we can safely ignore its contribution by cutting the corresponding
rows and columns from the starting Fisher matrix.}.

In Table~\ref{tab:fisher_scenarios} we define: the general Fisher matrix in each redshift bin, $\mathcal{F}^{\rm bin}$; its marginalization over the uninformative parameters\footnote{Note that the marginalization over the bias}, depending on the considered scenario, $\mathcal{F}^{\rm bin}_{\rm marg}$; the projection of $\mathcal{F}^{\rm bin}_{\rm marg}$ onto the relevant two-dimensional parameter space related to the interacting model $(\mathcal{F}^{\rm bin}_{\rm cosmo})$ in each redshift bin; and $\mathcal{F}^{\rm tot}_{\rm cosmo}$, the total Fisher onto these two parameters obtained summing up contribution from each redshift bin. The values reported in our Tables are derived from $\mathcal{F}^{\rm bin}_{\rm cosmo}$ and $\mathcal{F}^{\rm tot}_{\rm cosmo}$ after inversion, i.e., the errors on a given parameter are obtained from $\sigma_{p_i} = \sqrt{F^{-1}_{ii}}$.

In this analysis, we generate the transfer function with the Boltzmann code CLASS \citep{class}
adapted to our perturbation equations. Our fiducial cosmology is a flat $\Lambda$CDM model
with the following parameters: $\Omega_m = 0.31$, $h = 0.68$,
$n_s = 0.96$, and $\sigma_{8,0} = 0.82$.
As mentioned earlier, the surveys considered in this analysis are
J-PAS~\citep{Benitez:2014ibt,mini}, DESI~\citep{Aghamousa:2016zmz},
and \textit{Euclid} \citep{Euclid}. The surveys specifications used in our forecast
analysis can be found in Table~\ref{tab:surveys} and also in Refs.~\citep{antonio,EucFish}.
For completeness, all the quantities used to obtain the Fisher matrices elements are
shown in the Appendix A.

\section{Results}
\label{concludesec}

\begin{figure*}
\begin{center}
\includegraphics[width=0.95\textwidth]{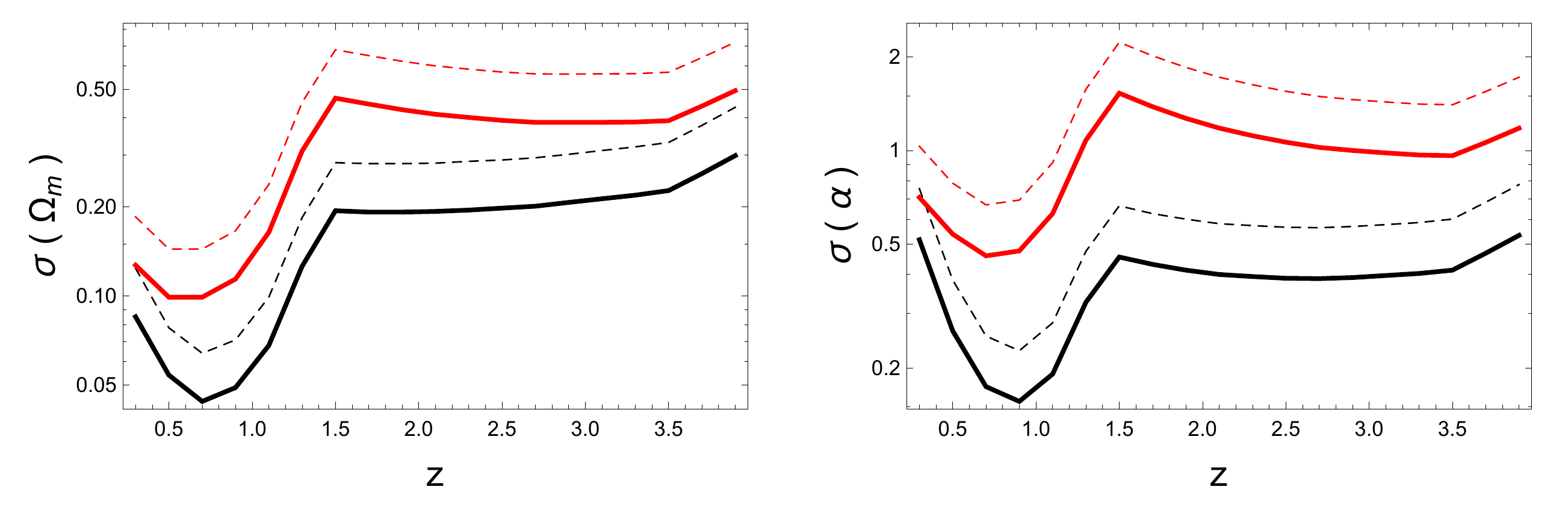}
\end{center}
\caption{Comparison of the J-PAS estimated errors for the interaction
parameter $\alpha$ and the matter density parameter $\Omega_{m}$
with the BAO+RSD data. The red and black lines correspond to the
case in which the baryonic and total matter growing functions are
used in the analysis, respectively.
Thin and thick lines are the plots for J-PAS 4000 deg$^2$ and
8500 deg$^2$, respectively.}
\label{bxm'}
\end{figure*}

In this section we summarize the main results of the present work. In Figs.~\ref{bxm'}-\ref{omega}, we plot the J-PAS forecasted errors
on $\Omega_m$ and $\alpha$ as functions of the redshift.
In Appendixes B, C, D, we present the errors at different redshifts
derived from the analysis of BAO, BAO+RSD, and BAO+RSD+PS, respectively.
In Tables III-XIV, we also show the errors found when we use $\delta_m$ instead of
$\delta_b$ in Eq.~(\ref{eq:power_v0}) for the sake of completeness and comparison,
together with those for DESI and Euclid.

In Figs.~\ref{bxm'} and \ref{bxm}, the errors on the parameters
$\Omega_m$ and $\alpha$ are larger when only baryonic tracers
are considered in most of the redshift range.
This is attributed to the fact that
$\delta_m$ is subject to stronger enhancement or suppression
in comparison to $\delta_b$ (see Fig.~\ref{ffig}), so the analysis
based on the total matter contrast gives rise to tighter bounds
on the coupling $\alpha$.
In Fig.~\ref{alpha1'} the difference between baryonic and total matter
is particularly clear for ELGs and LRGs at low redshifts, where
BAO and RSD are used with single tracers.
As we observe in Fig.~\ref{alpha1},
constraining the whole power spectrum leads to more accurate results
as compared to the case where only BAO and RSD are fitted.

\begin{figure*}
\begin{center}
\includegraphics[width=0.95\textwidth]{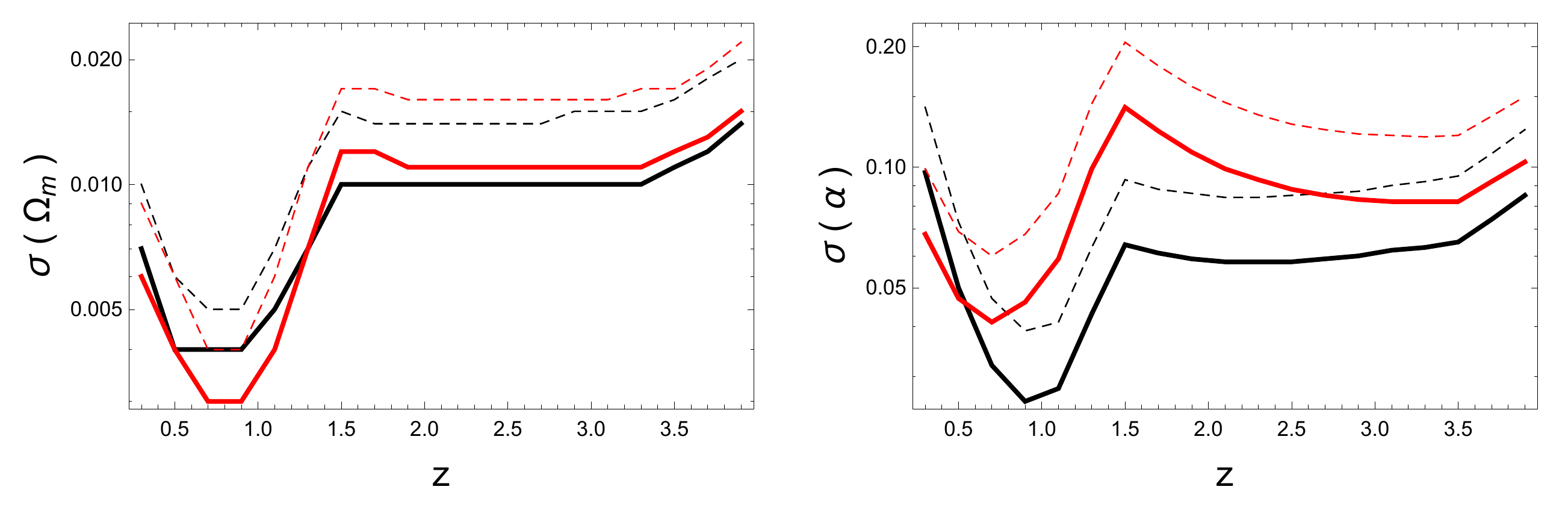}
\end{center}
\caption{The same as Fig.~\ref{bxm'}, but with the analysis
based on BAO + RSD + PS.}
\label{bxm}
\end{figure*}

\begin{figure*}
\begin{center}
\includegraphics[width=0.95\textwidth]{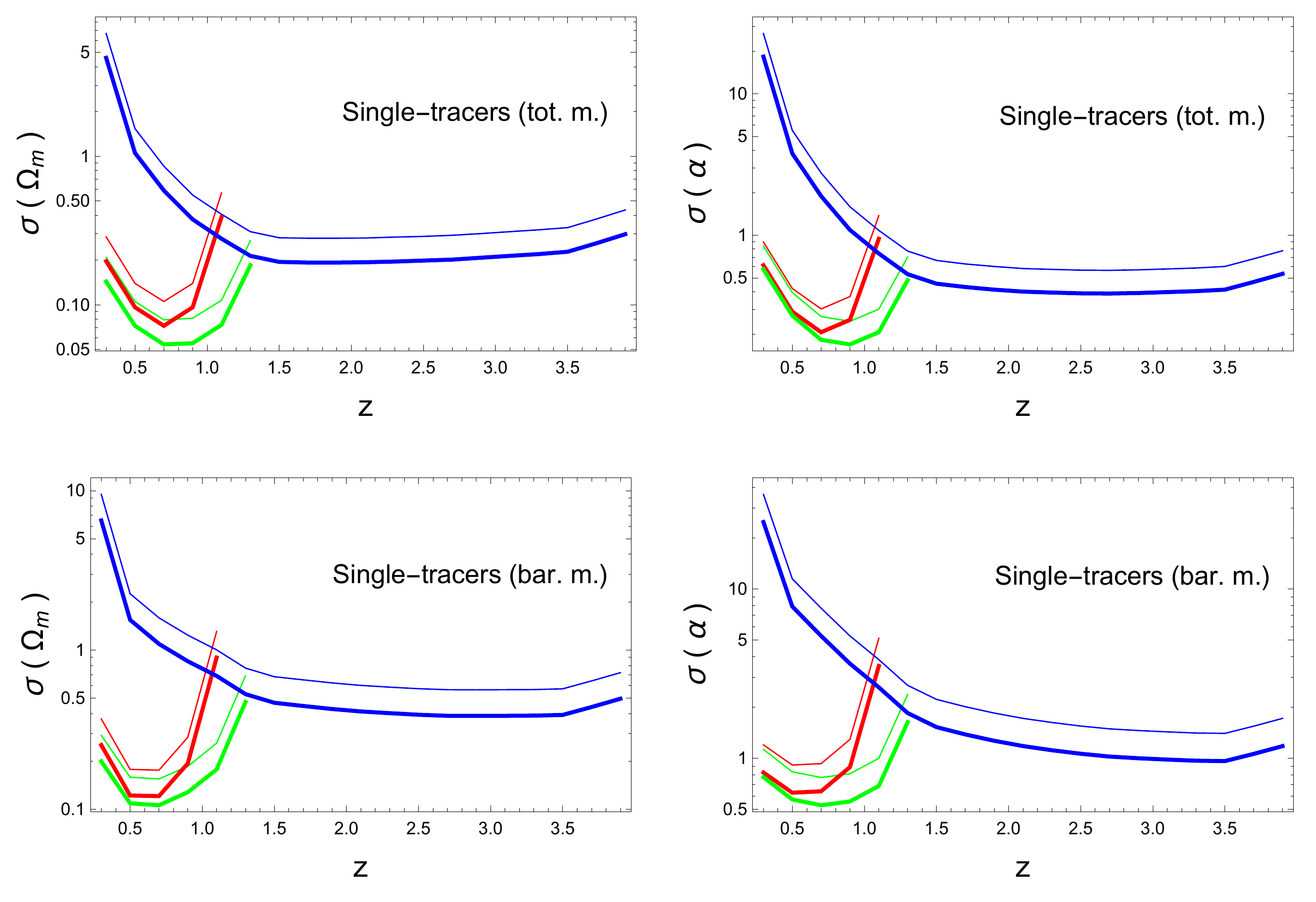}
\end{center}
\caption{Estimated error for $\Omega_{m}$ and $\alpha$
as a function of $z$ with the BAO+RSD data.
The estimations are from ELGs (green), LRGs (red) and QSOs (blue) power spectra
for J-PAS survey in both 4000 deg$^2$ (thin lines) and 8500 deg$^2$
(thick lines) regions.}
\label{alpha1'}
\end{figure*}

\begin{figure*}
\begin{center}
\includegraphics[width=0.95\textwidth]{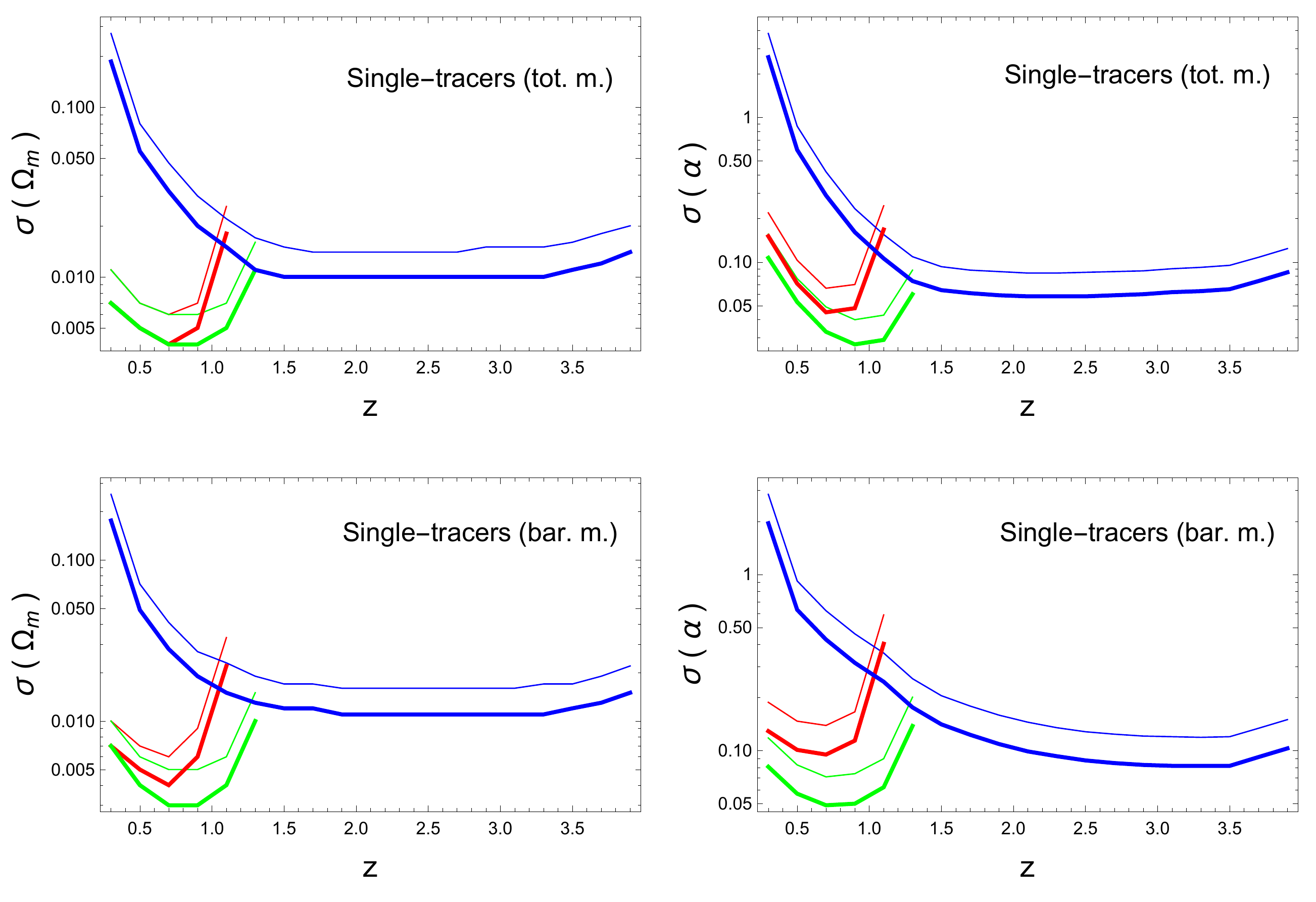}
\end{center}
\caption{The same as Fig.~\ref{alpha1'}, but with the analysis
based on BAO + RSD + PS.}
\label{alpha1}
\end{figure*}

\begin{figure*}
\begin{center}
\includegraphics[width=0.95\textwidth]{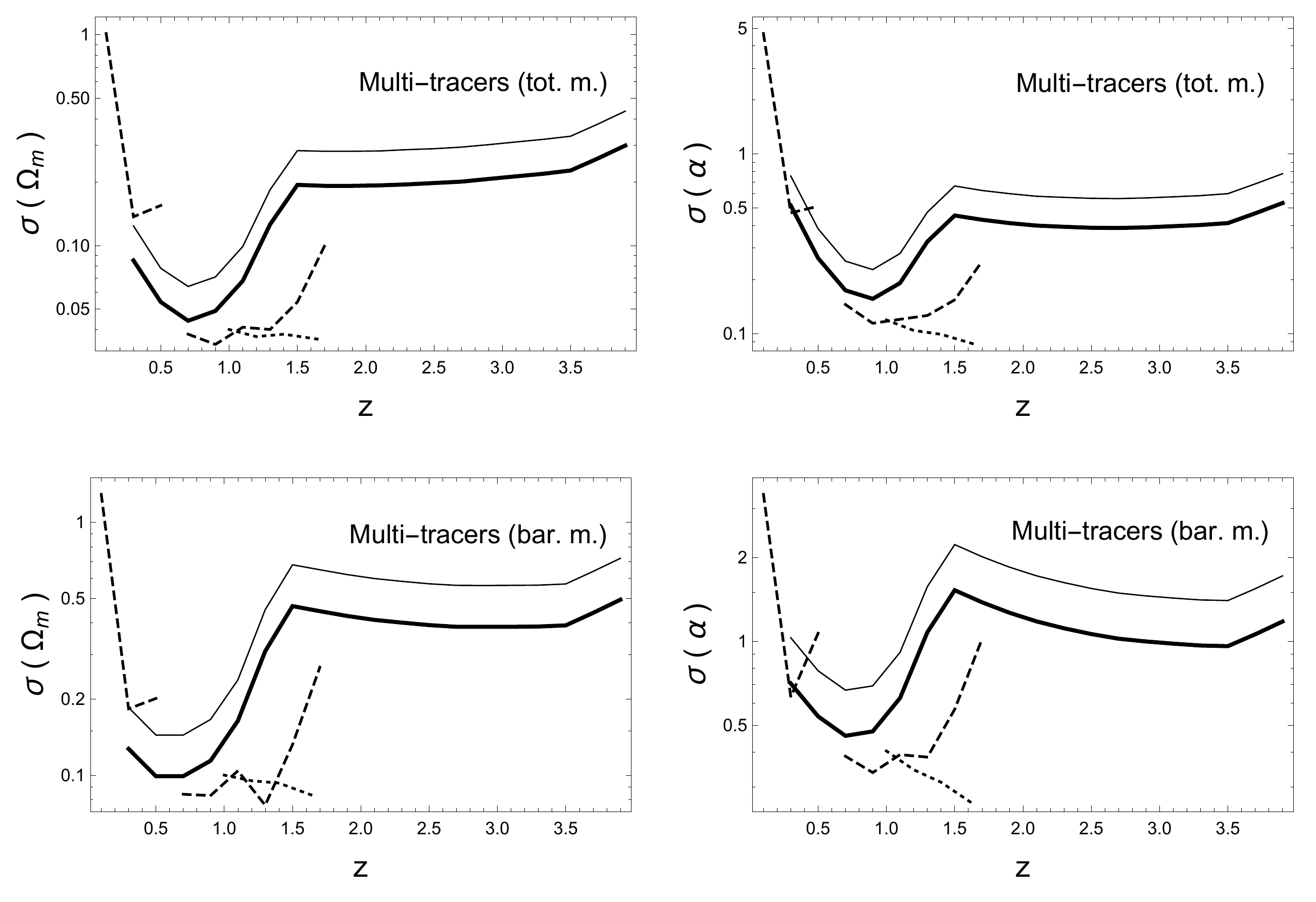}
\end{center}
\caption{Estimated error for $\Omega_{m}$ and $\alpha$
as a function of $z$ with the BAO+RSD data.
We show estimations from multi-tracers for Euclid (dotted),
DESI (dashed; BGS at lower redshifts; LRG+ELG+QSO at higher redshifts),
and for J-PAS survey in both 4000 deg$^2$ (thin solid) and
8500 deg$^2$ (thick solid) regions.}
\label{omega'}
\end{figure*}

\begin{figure*}
\begin{center}
\includegraphics[width=0.925\textwidth]{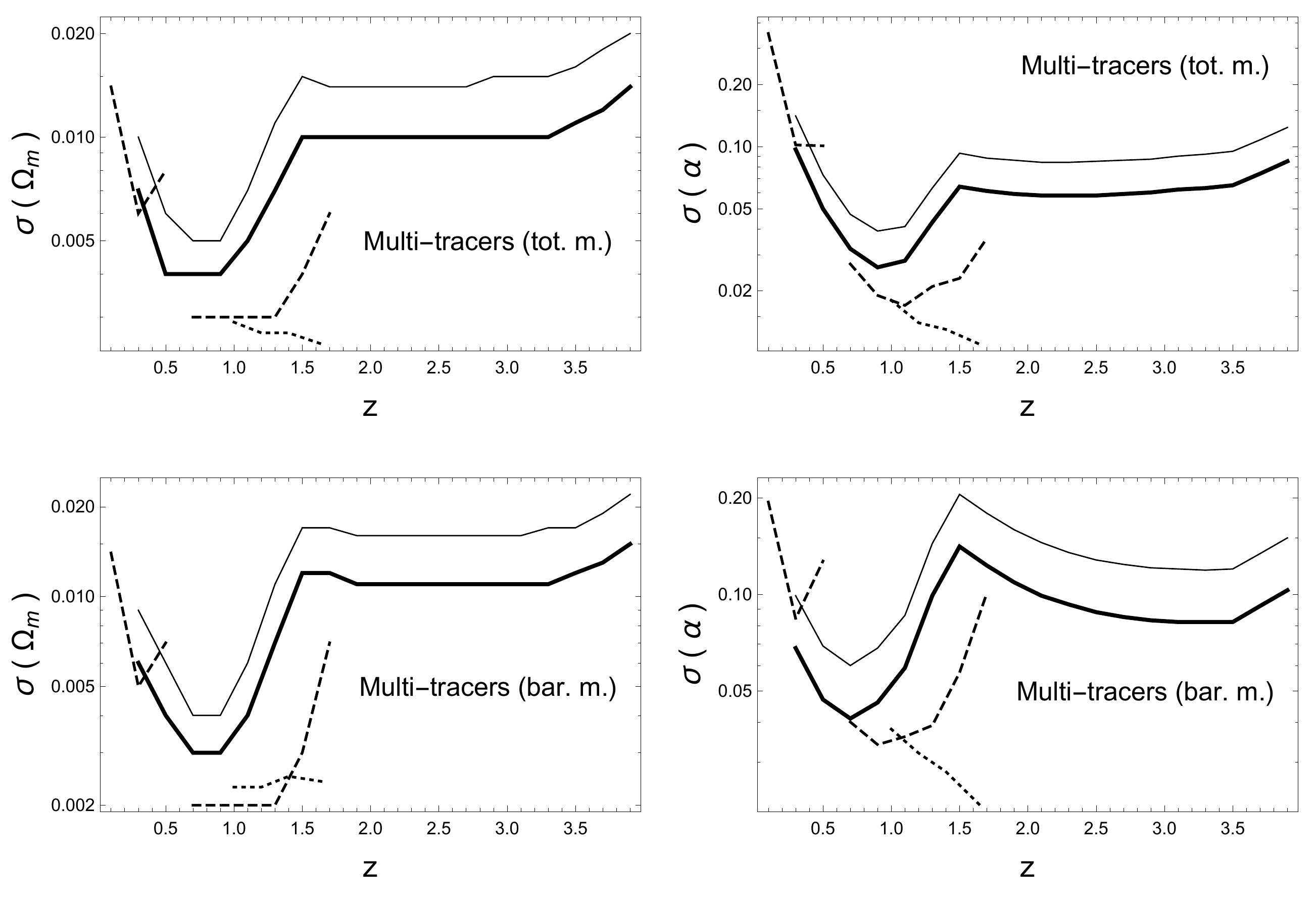}
\end{center}
\caption{The same as Fig.~\ref{omega'}, but with the analysis
based on BAO + RSD + PS.}
\label{omega}
\end{figure*}

In Figs.~\ref{omega'} and \ref{omega}, we plot the estimated errors for
$\Omega_{m}$ and $\alpha$ derived by using multi-tracers
for both total matter and baryons.
Besides J-PAS, we also show estimations of the DESI and
Euclid surveys for comparison.
We see from our results that J-PAS is competitive to DESI and Euclid for $z < 0.6$,
which confirms the results of a previous J-PAS forecast on non-interacting
dark energy models and modified gravity theories \citep{antonio}.

\begin{figure*}
\begin{center}
\includegraphics[width=0.95\textwidth]{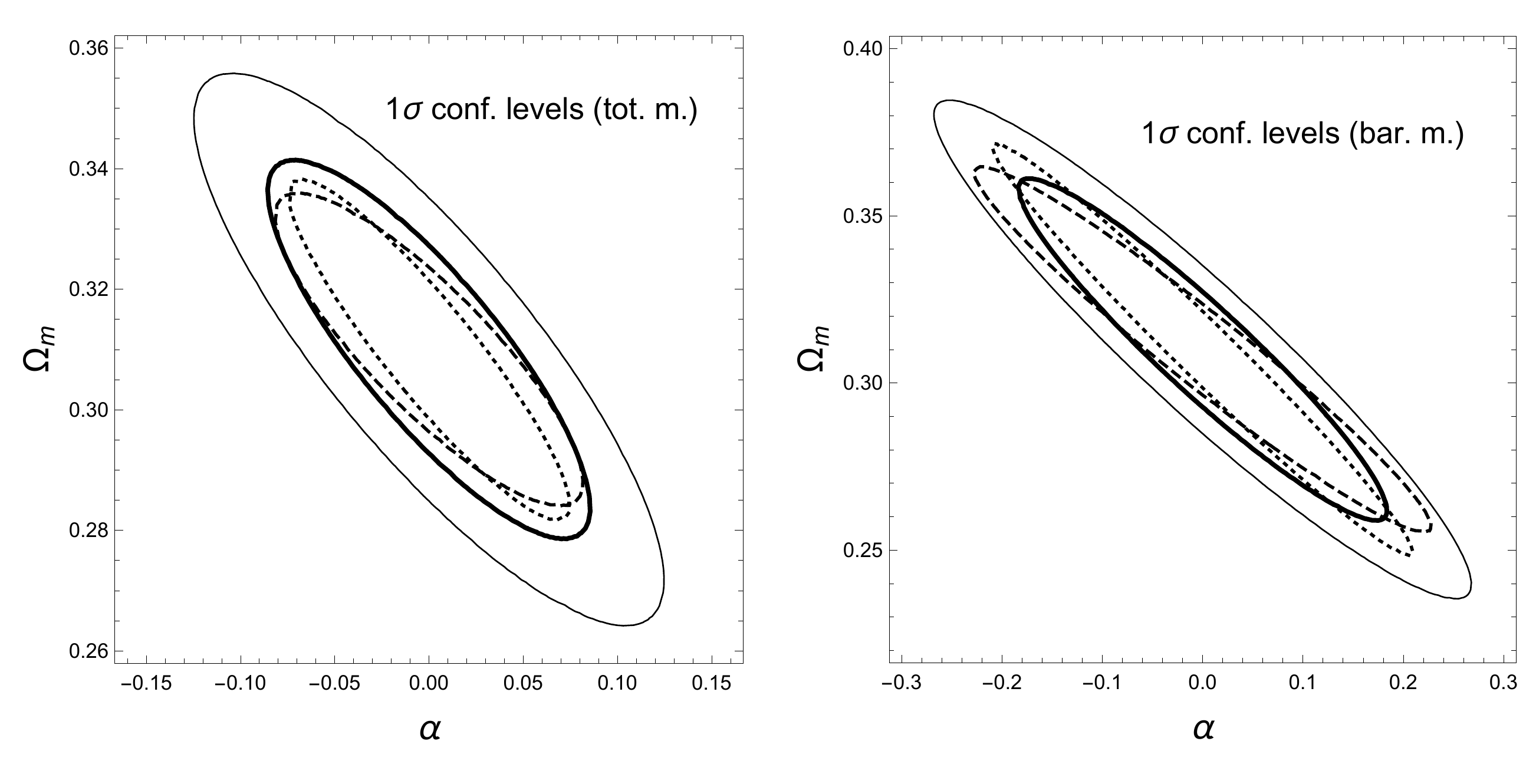}
\end{center}
\caption{1$\sigma$ contour error around $\alpha =0$ and $\Omega_{m}=0.31$
using multi-tracers for J-PAS 4000 deg$^2$ (thin solid), J-PAS 8500 deg$^2$ (thick solid),
DESI (dashed) and Euclid (dotted) surveys, with the BAO+RSD data.}
\label{alpha1b'}
\end{figure*}

\begin{figure*}
\begin{center}
\includegraphics[width=0.925\textwidth]{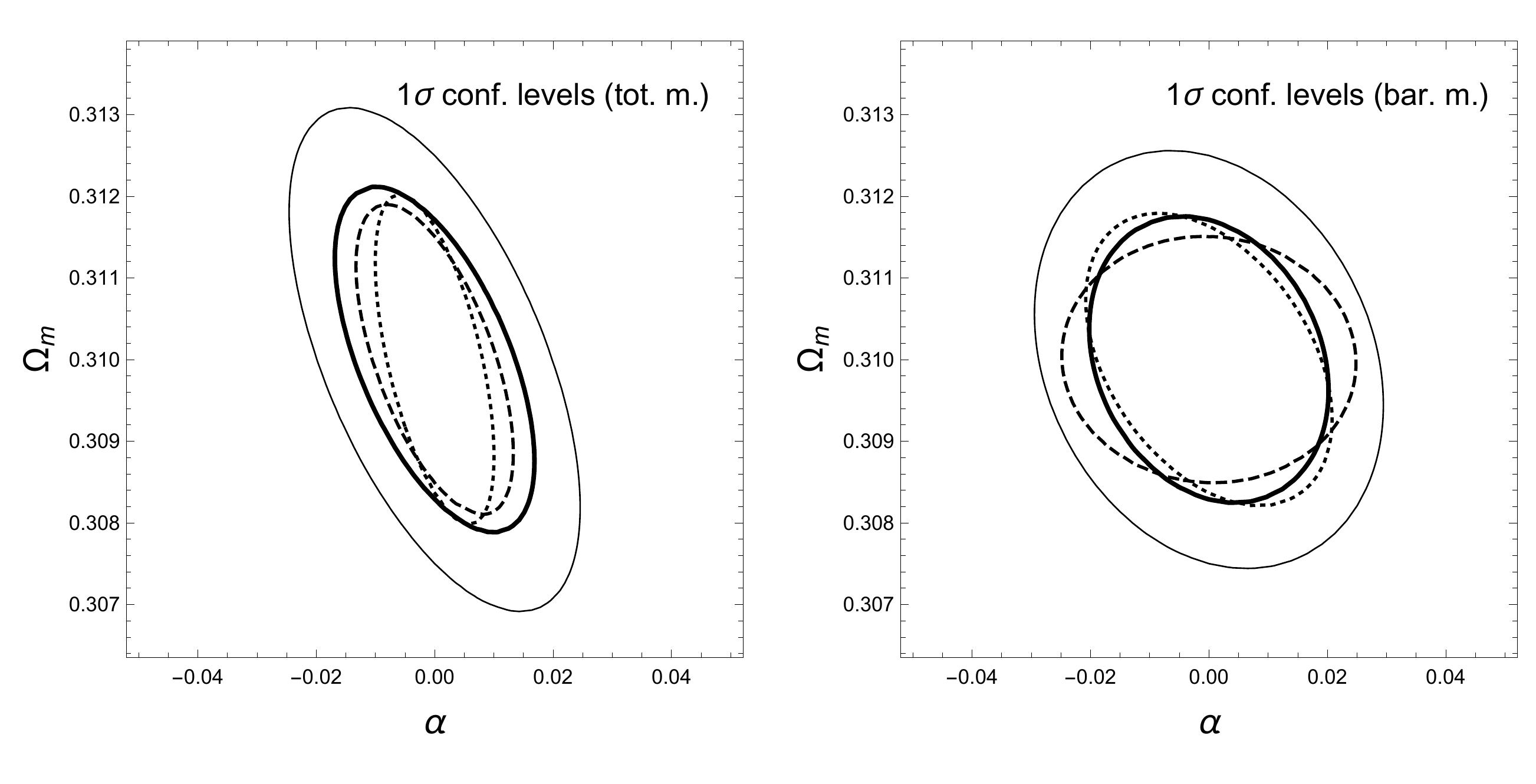}
\end{center}
\caption{The same as Fig.~\ref{alpha1b'}, but with the analysis
based on BAO + RSD + PS.}
\label{alpha1b}
\end{figure*}

Finally, we show in Figs.~\ref{alpha1b'} and \ref{alpha1b} the confidence regions
in the $\{\Omega_{m}, \alpha\}$ plane for BAO+RSD and BAO+RSD+PS, respectively.
We see that J-PAS with $8500$ deg$^2$ is very competitive to the other two surveys considered in our analysis.
At low redshifts ($z < 0.6$), when the presence of dark energy
and the dark sector interaction is more relevant, J-PAS provides the best constraints on the parameters thanks to its high galaxy densities detectable in that redshift range. In the BAO+RSD+PS case, the J-PAS predicted
errors are comparable to the actual errors derived from JLA SN Ia + Planck CMB data \citep{micol}. The full tables, showing all the results derived in our analysis, are shown in Appendices B, C and D.

\section{Conclusions}

The goal of the analysis performed in this paper is to report the J-PAS forecasts for the parameter estimation
in a class of interacting vacuum energy models.
The vacuum energy does not give rise to an additional dynamical
degree of freedom, but the interaction between vacuum energy
and CDM modifies the background cosmological dynamics
through an energy exchange.
At the level of perturbations, there are in general contributions
to the interacting four vector $Q^{\mu}$ arising from
energy and momentum transfers, which are weighed by
$Q \mu^{\mu}$ and $q^{\mu}$ respectively.
We focused on the case where the momentum transfer is absent ($q^{\mu}=0$),
under which the four velocities of vacuum energy and CDM are
equivalent to each other.
Then the vacuum energy perturbation $\delta \Lambda$ is
related to the CDM four velocity potential $\theta$ as
$\delta \Lambda=Q \theta$, so the system of
perturbation equations of motion is closed.

The interacting vacuum energy is chosen to be of the form (\ref{asu})
with $\sigma$ given by (\ref{sigma}), in which case $Q=-\alpha \sigma H^{-2\alpha-1}\rho_m$.
Then, the Hubble parameter is analytically known as a function of
the redshift, see Eq.~(\ref{H}).
The coupling constant $\alpha$ is the only additional parameter
to those appearing in the $\Lambda$CDM model ($\alpha=0$).
For $\alpha>0$, the growth rate of CDM density contrast $\delta_c$ is
enhanced in comparison to the $\Lambda$CDM.
This also leads to the larger growth rates of baryon and total
matter density contrasts. For $\alpha<0$, the evolution of
$\delta_c$, $\delta_b$, and $\delta_m$
is suppressed relative to the $\alpha=0$ case.
These properties manifest themselves in the observations
of the galaxy power spectrum including BAO and RSD.

We carried out the J-PAS forecasts for the matter density parameter
$\Omega_m$ and the coupling $\alpha$.
J-PAS will cover a region of $8500$ deg$^2$ in the northern sky.
We extended the analysis for a smaller covering of $4000$ deg$^2$ with
a more conservative estimation of the impact of the data.
For a $8500$ deg$^2$ covering, it has been shown that J-PAS will lead to
parameter estimations comparable to those obtained from DESI and
Euclid for the same class of interacting models.
In particular, for the redshift range $z < 0.6$, the error bars in J-PAS measurements
can be even smaller than those provided by DESI and
Euclid -- see Figs.~\ref{omega'} and \ref{omega}.
In general, the J-PAS constraints will tighten
considerably the estimations of those parameters with respect to previous
large-scale structure data, with precision comparable to that
we have with actual CMB data. We conclude that confirming or ruling out the $\Lambda$CDM model by finding an interaction
signature in a joint analysis of SN Ia, CMB, and large-scale structure data,
with any of the three surveys considered in our analysis or
by combining them, is a real and promising possibility.

\section*{Acknowledgements}

We are thankful to our colleagues of J-PAS Theory Working Group for helpful discussions. MB acknowledges Istituto Nazionale di Fisica Nucleare (INFN), sezione di Napoli, iniziativa specifica QGSKY. RvM acknowledges support from the Programa de Capacita\c{c}\~{a}o Institucional do Observatorio Nacional PCI/ON/MCTI. SC is supported by CNPq (Grants No. 307467/2017-1 and 420641/2018-1). JA is supported by CNPq (Grants No. 310790/2014-0 and 400471/2014-0) and FAPERJ (Grant No. 233906). JCF is supported by CNPq (Grant No. 304521/2015-9) and FAPES (Grant No. 78/2017). ST is supported by the Grant-in-Aid for Scientific Research Fund of the JSPS No.\,19K03854. VM thanks CNPq (Brazil) and FAPES (Brazil) for partial financial support. This project has received funding from the European Union’s Horizon 2020 research and innovation programme under the Marie Skłodowska-Curie grant agreement No 888258.

This paper has gone through internal review by the J-PAS collaboration. Funding for the J-PAS Project has been provided by the Governments of Spain and Arag\'on through the Fondo de Inversi\'on de Teruel, European FEDER funding and the MINECO and by the Brazilian agencies FINEP, FAPESP, FAPERJ and by the National Observatory of Brazil.

\bibliographystyle{apsrev4-1}
\bibliography{biblio}

\appendix

\section{Fisher Matrices derivatives}

The elements of the Fisher Matrices are better visualized if we rewrite
the observed matter power spectrum considering the mapping between
the fiducial cosmology and the real unknown cosmological background:
\begin{align}
\mathcal{P}_{\rm obs}(k^{\rm fid}_{\perp},k^{\rm fid}_{||},z)
&= \alpha^{2}_{\perp} \alpha_{||} \Bigg[ b_{s}(z)+ f_{s}(z)
\left( \frac{k^{2,{\rm fid}}_{||} \alpha^{2}_{||}}{k^{2,{\rm fid}}_{||}
\alpha^{2}_{||} + k^{2,{\rm fid}}_{\perp} \alpha^{2}_{\perp}}\right) \Bigg]^{2}
\frac{\mathcal{P}_{L,0}(k=\sqrt{k^{2,{\rm fid}}_{||} \alpha^{2}_{||}
+ k^{2,{\rm fid}}_{\perp} \alpha^{2}_{\perp}})}{\sigma^{2}_{8,0}}
\exp\left(-k^{2} \mu^2 \frac{\sigma^{2}_{z}}{\alpha^{2}_{||}
H^{2,{\rm fid}}}\right) + \mathcal{P}_{{\rm shot}},
\end{align}
where, following BAO literature, we have defined the tangential and radial BAO modes as
\begin{align}
\label{eq:BAO-modes}
\alpha_{\perp}(z) &\equiv \frac{r^{\rm fid}_{\perp}(z)}{r_{\perp}(z)}
= \frac{D^{{\rm fid}}_{A}(z)}{D_{A}(z)}\, , \\
\alpha_{||}(z) &\equiv \frac{r^{\rm fid}_{||}(z)}{r_{||}(z)}
= \frac{H(z)}{H^{\rm fid}(z)}\; ,
\end{align}
where
\begin{align}
r_{||}(z) &= \frac{c}{H(z)\,r_{s}(z_{d})}\, , \\ r_{\perp}(z) &= \frac{D_{A}(z)}{r_{s}(z_{d})}\, ,
\end{align}
where we explicitly wrote the speed of light c. The angular diameter distance is related to the comoving distance defined in Eq.~(\ref{chi}) by $D_{A}(z) \equiv \chi / (1+z)$ and $r_s$ is the sound horizon at dragging epoch, $z_d$. Note that we are interested in errors on $\ln D_{A}$ and $\ln H$ which, by previous definitions,
are given by
\begin{align}
\frac{\sigma_{\alpha_{\perp}}}{\alpha_{\perp}} &= \frac{\sigma_{D_{A}}}{D_{A}} = \sigma_{\ln D_{A}}, \\ \frac{\sigma_{\alpha_{||}}}{\alpha_{||}} &= \frac{\sigma_{H}}{H} = \sigma_{\ln H}\,,
\end{align}
and considering that the Fisher Matrices have to be evaluated at the fiducial cosmology,
for which we have $\alpha_{\perp} = 1$ and $\alpha_{||} = 1$.
We have also used the definitions
\begin{align}
k^{2} &= k^{2}_{||} + k^{2}_{\perp}, \\
\mu^{2} &= \frac{k^{2}_{||}}{k^{2}}\, ,
\end{align}
and the transformation rules
\begin{align}
k_{\perp} &= k^{\rm fid}_{\perp} \alpha_{\perp}, \\
k_{||} &= k^{\rm fid}_{||} \alpha_{||}.
\end{align}
In the following, we report the expressions of all the functions
which are needed to calculate the Fisher matrices:
\begin{itemize}
\item Derivatives of the power spectrum, $\ln \mathcal{P}_{\rm obs}$,
with respect to $\alpha_{\perp}$:
\begin{align}
\frac{{\rm d} \ln \mathcal{P}_{\rm obs}}{{\rm d} \alpha_{\perp}}= \left[ \frac{2}{\alpha_{\perp}} + \frac{2 f_{s}}{b_{s}+f_{s} \mu^{2}}
\frac{{\rm d} \mu^2}{{\rm d} \alpha_{\perp}} + \frac{{\rm d} \ln \mathcal{P}_{L,0}}
{{\rm d} k} \frac{{\rm d} k}{{\rm d} \alpha_{\perp}}
- \frac{k \sigma^{2}_{z}}{\alpha_{||}^2 H^{2,{\rm fid}}}\left( 2 \mu^2
\frac{\partial k}{\partial \alpha_{\perp}} + k \frac{\partial \mu^2}
{\partial \alpha_{\perp}} \right)\right] \frac{\mathcal{P}_{g}}{\mathcal{P}_{\rm obs}}\,,
\end{align}
with
\begin{align}
\frac{{\rm d} \mu^2}{{\rm d} \alpha_{\perp}}
&= -\frac{2}{\alpha_{\perp}}\mu^{2}(1-\mu^{2})\,, \\
\frac{{\rm d} k}{{\rm d} \alpha_{\perp}} &= \frac{k}{\alpha_{\perp}}(1-\mu^{2})\,.
\end{align}
\item Derivatives of the power spectrum, $\ln \mathcal{P}_{\rm obs}$,
with respect to $\alpha_{||}$, when we consider the baryonic matter
linear contrast, $\delta_b$:
\begin{align}
\frac{{\rm d} \ln \mathcal{P}_{\rm obs}}{{\rm d} \alpha_{||}}= \left[ \frac{1}{\alpha_{||}} + \frac{2 f_{s}}{b_{s}+f_{s} \mu^{2}}
\frac{{\rm d} \mu^2}{{\rm d} \alpha_{||}} + \frac{{\rm d} \ln \mathcal{P}_{L,0}}{{\rm d} k}
\frac{{\rm d} k}{{\rm d} \alpha_{||}}
+ \frac{k \sigma^{2}_{z}}{\alpha_{||}^2 H^{2,{\rm fid}}} \left( -2 \mu^2 \frac{\partial k}{\partial \alpha_{||}} - k \frac{\partial \mu^2}{\partial \alpha_{||}}
+\frac{2 k \mu^2}{\alpha_{||}}\right) \right] \frac{\mathcal{P}_{g}}{\mathcal{P}_{\rm obs}}\,,
\end{align}
and when we consider the total (dark plus baryonic) matter linear contrast, $\delta_m$:
\begin{align}
\frac{{\rm d} \ln \mathcal{P}_{\rm obs}}{{\rm d} \alpha_{||}} =
\left[ \frac{1}{\alpha_{||}} + \frac{2}{b_{s}+f_{s} \mu^{2}} \left(f_{s} \frac{{\rm d} \mu^2}{{\rm d} \alpha_{||}}
+ \frac{\partial f_{s}}{\partial \alpha_{||}} \mu^2 \right)
+ \frac{{\rm d} \ln \mathcal{P}_{L,0}}{{\rm d} k}
\frac{{\rm d} k}{{\rm d} \alpha_{||}} - \frac{k \sigma^{2}_{z}}{\alpha_{||}^2 H^{2,{\rm fid}}}
\left( 2 \mu^2 \frac{\partial k}{\partial \alpha_{||}} + k \frac{\partial \mu^2}
{\partial \alpha_{||}} -\frac{2 k \mu^2}{\alpha_{||}}\right) \right] \frac{\mathcal{P}_{g}}{\mathcal{P}_{\rm obs}}\,,
\end{align}
with
\begin{align}
\frac{{\rm d} \mu^2}{{\rm d} \alpha_{||}} &=
\frac{2}{\alpha_{||}}\mu^{2}(1-\mu^{2})\,,\\
\frac{{\rm d} k}{{\rm d} \alpha_{||}} &= \frac{k}{\alpha_{||}}\mu^{2}\,.
\end{align}
The growth rate of total matter relevant to the RSD measurements
is given by $f_s=(f_m+g_m)\sigma_8$, where
\begin{align}
g_m \equiv \frac{Q}{H \rho_m} = \frac{\Gamma}{H} = -3\alpha(1-\Omega_m)
\left( \frac{H_0}{H} \right)^{2(\alpha+1)}
= -3\alpha (1-\Omega_{m}) \left(\frac{H_{0}}{\alpha_{||}H^{\rm fid}}\right)^{2(\alpha+1)}\,.
\end{align}
Then, we have
\begin{align}
\frac{\partial f_{s}}{\partial \alpha_{||}} &= \frac{\partial g_m}{\partial \alpha_{||}} \sigma_{8}\; , \\
\frac{\partial g_m}{\partial \alpha_{||}} &= 3\alpha(1-\Omega_m) \frac{2(\alpha+1)}{\alpha^{2\alpha+3}_{||}}
\left(\frac{H_0}{H^{\rm fid}}\right)^{2(\alpha+1)}\,.
\end{align}
\item Derivatives of the power spectrum, $\ln \mathcal{P}_{\rm obs}$,
with respect to the growth rate, $f_{s}$:
\begin{equation}
\frac{{\rm d} \ln \mathcal{P}_{\rm obs}}{{\rm d} f_{s}}
= \frac{2 \mu^{2}}{b_{s}+ f_{s}\mu^{2}} \frac{\mathcal{P}_{g}}{\mathcal{P}_{\rm obs}}\,.
\end{equation}
\item Derivatives of the power spectrum, $\ln \mathcal{P}_{\rm obs}$,
with respect to the galaxy bias, $b_{s}$:
\begin{equation}
\frac{{\rm d} \ln \mathcal{P}_{\rm obs}}{{\rm d} b_{s}}
= \frac{2}{b_{s}+ f_{s}\mu^{2}} \frac{\mathcal{P}_{g}}{\mathcal{P}_{\rm obs}}\,.
\end{equation}
\item Derivatives of the power spectrum, $\ln \mathcal{P}_{\rm obs}$,
with respect to the shot noise, $\mathcal{P}_{\rm shot}$:
\begin{equation}
\frac{{\rm d} \ln \mathcal{P}_{\rm obs}}{{\rm d} \mathcal{P}_{\rm shot}}
= \frac{1}{\mathcal{P}_{\rm obs}}\,,
\end{equation}
and we have assumed $\mathcal{P}^{\rm fid}_{\rm shot} = 0$ for the fiducial model.
\item Derivatives of the power spectrum, $\ln \mathcal{P}_{\rm obs}$,
with respect to $\Omega_m$:
\begin{align}
\frac{\partial \ln \mathcal{P}_{\rm obs}}{\partial \Omega_m}
&= \frac{\mathcal{P}_{g}}{\mathcal{P}_{\rm obs}} \left[ \frac{2}{\alpha_{\perp}} \frac{\partial \alpha_{\perp}}{\partial \Omega_m} +
\frac{1}{\alpha_{||}} \frac{\partial \alpha_{||}}{\partial \Omega_m}
+\frac{2}{b_{s} + f_{s} \mu^2} \left(  \frac{\partial b_{s}}{\partial \Omega_m} + \frac{\partial f_{s}}{\partial \Omega_m} \mu^2 + f_{s}  \frac{\partial \mu^2}{\partial \Omega_m} \right)
+\frac{1}{\mathcal{P}_{L,0}}  \frac{\partial \mathcal{P}_{L,0}}{\partial \Omega_m} \right. \nonumber \\
&- \left.\frac{k \sigma^{2}_{z}}{\alpha_{||}^2 H^{2,{\rm fid}}}
\left( 2 \mu^2 \frac{\partial k}{\partial \Omega_m}
+ k \frac{\partial \mu^2}{\partial \Omega_m}
- \frac{2k \mu^2}{\alpha_{||}} \frac{\partial \alpha_{||}}{\partial \Omega_m}\right) \right]\,,
\end{align}
with
\begin{align}
\frac{\partial k}{\partial \Omega_m}&= \frac{\partial k}{\partial \alpha_{\perp}}
\frac{\partial \alpha_{\perp}}{\partial \Omega_m} + \frac{\partial k}{\partial \alpha_{||}}
\frac{\partial \alpha_{||}}{\partial \Omega_m}\, , \\
\frac{\partial \mu^2}{\partial \Omega_m} &=  \frac{\partial \mu^2}{\partial \alpha_{\perp}} \frac{\partial \alpha_{\perp}}{\partial \Omega_m} + \frac{\partial \mu^2}{\partial \alpha_{||}} \frac{\partial \alpha_{||}}{\partial \Omega_m}.
\end{align}
\item Derivatives of the radial mode, $\alpha_{||}$, with respect to $\Omega_m$:
\begin{align}
\frac{\partial \alpha_{||}}{\partial \Omega_{m}} &= \frac{1}{E^{\rm fid}} \frac{\partial E}{\partial \Omega_{m}}\, , \\
\frac{\partial E}{\partial \Omega_{m}} &= \frac{\left[ a^{-3(\alpha+1)}-1\right]}{2 (\alpha +1)}\frac{\left[\Omega_m \left(a^{-3 (\alpha+1)}-1\right)+1
\right]^{-\alpha/(\alpha+1)}}{\sqrt{\left[
\Omega_m \left(a^{-3(\alpha+1)}-1\right)+1\right]^{1/(\alpha+1)}}}\,,
\end{align}
where $E \equiv H/H_0$.
\item Derivatives of the tangential mode, $\alpha_{\perp}$,
with respect to $\Omega_m$:
\begin{align}
\frac{\partial \alpha_{\perp}}{\partial \Omega_{m}} &=
- \alpha_{\perp} \frac{1}{d_{C}} \frac{\partial d_{C}}{\partial \Omega_{m}}\, , \\
d_{C}(z) &= \int_{0}^{z} {\rm d}z' \frac{1}{E(z')}\,, \\
\frac{\partial d_{C}}{\partial \Omega_{m}} &= - \int^{z}_{0} {\rm d}z'
\frac{1}{E^{2}(z')} \frac{\partial E(z')}{\partial \Omega_{m}}\,.
\end{align}
\item Derivatives of the linear power spectrum, $\mathcal{P}_{L,0}$,
with respect to $\Omega_m$:
\begin{equation}
\frac{\partial \mathcal{P}_{L,0}}{\partial \Omega_m} =
\mathcal{P}_{0} n_{s} k^{n_s-1} \left( \frac{\partial k}{\partial \Omega_m}
\right) \mathcal{T}^{2} + \mathcal{P}_{0} k^{n_s}
\frac{\partial \mathcal{T}^2}{\partial \Omega_m}\,,
\end{equation}
where derivatives of the transfer function are calculated numerically.
\item Derivatives of the growth rate, $f_{s}=f_{b}\sigma_8$,
with respect to $\Omega_m$: when considering only baryonic matter,
\begin{equation}
\frac{\partial f_{s}}{\partial \Omega_m} =
\left( \frac{\partial f_b}{\partial \Omega_m}\right) \sigma_8
+ f_b \left( \frac{\partial \sigma_8}{\partial \Omega_m} \right)\,.
\end{equation}
For total matter, the growth rate relevant to the RSD measurements
is given by $f_s=(f_m+g_m)\sigma_8$, where $g_m=Q/(H \rho_m)$.
Then, it follows that
\begin{equation}
\frac{\partial f_{s}}{\partial \Omega_m} = \left( \frac{\partial f_m}
{\partial \Omega_m}+\frac{\partial g_m}{\partial \Omega_m}\right) \sigma_8
+ (f_m+g_m) \left( \frac{\partial \sigma_8}{\partial \Omega_m} \right)\,,
\end{equation}
where
\begin{align}
\frac{\partial g_m}{\partial \Omega_m} &=
3\alpha \left( \frac{H_0}{\alpha_{||}H^{\rm fid}}\right)^{2(\alpha+1)}
\left[1+\frac{2(1-\Omega_m)(1+\alpha)}{\alpha_{||}} \frac{\partial \alpha_{||}}{\partial \Omega_m}\right]\, ,
\end{align}
and
\begin{equation}
\frac{\partial \sigma_{8}}{\partial \Omega_m} =
\frac{\sigma_{8,0}}{\delta_{m,0}} \left( \frac{\partial \delta_{m}}
{\partial \Omega_m}\right) + \sigma_{8,0} \left(-\frac{\delta_{m}}
{\delta_{m,0}^2}\right) \left( \frac{\partial \delta_{m,0}}{\partial \Omega_m} \right) \, .
\end{equation}
Note that derivatives of the density contrasts $\delta_{b,m}$ and of
the growth rates $f_{b,m}$ are calculated numerically.
\item Derivatives of the bias factor, $b_{s}$, with respect to $\Omega_m$:
\begin{equation}
\frac{\partial b_{s}}{\partial \Omega_m}
= \left( \frac{\partial b_g}{\partial \Omega_m}\right) \sigma_8 +
b_g \left( \frac{\partial \sigma_8}{\partial \Omega_m} \right)\,.
\end{equation}
Given the different definition of the galaxy bias for the tracers
we have used in this work, in order to treat each of them in the same way,
we have set all $\partial b_g/\partial \Omega_m=0$.
\item
Derivatives of the power spectrum, $\ln \mathcal{P}_{\rm obs}$, with respect to $\alpha$:
the same equations hold, but with
\begin{align}
\frac{\partial E}{\partial \alpha} &= -\frac{\sqrt{\left[ \Omega_m \left(a^{-3(\alpha+1)}-1\right)+1\right]^{\frac{1}{\alpha +1}}}}{2 (\alpha +1)^2}
\left[\frac{3 (\alpha +1) \Omega_m a^{-3(\alpha+1)} \log (a)}{\Omega_m \left(a^{-3(\alpha+1)}-1\right)+1}
+\log \left\{ \Omega_m \left(a^{-3(\alpha+1)}-1\right)+1\right\}\right]\,, \\
\frac{\partial g_m}{\partial \alpha} &= -3(1-\Omega_m) \left( \frac{H_0}{\alpha_{||}H^{\rm fid}}\right)^{2(\alpha+1)}
\left[1+2\alpha \ln \left( \frac{H_0}{\alpha_{||}H^{\rm fid}} \right)-\frac{2\alpha(\alpha+1)}{\alpha_{||}}\frac{\partial \alpha_{||}}
{\partial \alpha}\right]\,.
\end{align}
\item Derivatives of the power spectrum, $\ln \mathcal{P}_{\rm obs}$, with respect to $h$:
the same equations hold, except for:
\begin{align}
\frac{\partial \alpha_{||}}{\partial h}&= \frac{\alpha_{||}}{h}\,, \\
\frac{\partial \alpha_{\perp}}{\partial h} &= \frac{\alpha_{\perp}}{h}\,, \\
\frac{\partial \delta_{b,m}}{\partial h} &=
\frac{\partial \sigma_{8}}{\partial h} = \frac{\partial f_{b,m}}{\partial h} =
\frac{\partial g_m}{\partial h} = 0\,.
\end{align}
\item Derivatives of the power spectrum, $\ln \mathcal{P}_{\rm obs}$, with respect to $n_s$:
all derivatives are zero, except for:
\begin{align}
\frac{\partial \mathcal{P}_{L,0}}{\partial n_s} &= \mathcal{P}_{L,0} \ln k\, , \\
\frac{\partial \ln \mathcal{P}_{\rm obs}}
{\partial n_s} &= \frac{\mathcal{P}_{g}}{\mathcal{P}_{\rm obs}}\ln k\; .
\end{align}
\end{itemize}

\clearpage

\section{Tables for BAO}
\label{AppenB}

{\renewcommand{\tabcolsep}{1.5mm}
{\renewcommand{\arraystretch}{1.8}
\begin{table}[hbt!]
\begin{minipage}{0.5\columnwidth}
\caption{Errors for \textit{Euclid} using baryons and total matter. As we are focusing only on geometrical cosmological quantities, there is no difference between the baryons and the total matter scenario in this case.}\label{tab:results_EUCLID-BAO}
\begin{center}
\resizebox{0.4\columnwidth}{!}{
\begin{tabular}{c|cc}
\hline
\hline
\multicolumn{3}{c}{BAO} \\
\hline
 & \multicolumn{2}{c}{ELG}\\
\hline
$z$ & $\sigma_{\Omega_m}$ & $\sigma_{\alpha}$ \\
\hline
\hline
1.00 & 0.128 & 0.503 \\
1.20 & 0.105 & 0.376 \\
1.40 & 0.097 & 0.321 \\
1.65 & 0.085 & 0.260 \\
\hline
Total & 0.044 & 0.148 \\
\hline
\hline
\end{tabular}}
\end{center}
\end{minipage}
\end{table}}}

{\renewcommand{\tabcolsep}{1.5mm}
{\renewcommand{\arraystretch}{1.8}
\begin{table}[hbt!]
\begin{minipage}{0.5\columnwidth}
\caption{Errors for DESI using baryons and total matter. As we are focusing only on geometrical cosmological quantities, there is no difference between the baryons and the total matter scenario in this case.}\label{tab:results_DESI-BAO}
\begin{center}
\resizebox{0.4\columnwidth}{!}{
\begin{tabular}{c|cc}
\hline
\hline
\multicolumn{3}{c}{BAO} \\
\hline
      & \multicolumn{2}{c}{Multi-tracers} \\
\hline
  $z$ & $\sigma_{\Omega_m}$ & $\sigma_{\alpha}$  \\
\hline
\hline
0.1 & 6.101 & 163.976 \\
0.3 & 0.804 & 7.875 \\
0.5 & 0.893 & 5.533 \\
0.7 & 0.160 & 0.754 \\
0.9 & 0.114 & 0.461 \\
1.1 & 0.123 & 0.455 \\
1.3 & 0.191 & 0.944 \\
1.5 & 0.222 & 0.920 \\
1.7 & 0.356 & 1.302 \\
\hline
Total & 0.051 & 0.211 \\
\hline
\hline
\end{tabular}}
\end{center}
\end{minipage}
\end{table}}}

{\renewcommand{\tabcolsep}{1.5mm}
{\renewcommand{\arraystretch}{1.8}
\begin{table*}
\begin{center}
\begin{minipage}{0.75\textwidth}
\caption{Errors for J-PAS $4000$ deg$^{2}$ using baryons and total matter. As we are focusing only on geometrical cosmological quantities, there is no difference between the baryons and the total matter scenario in this case.} \label{tab:results_JPAS4000-BAO}
\resizebox*{0.75\textwidth}{!}{
\begin{tabular}{c|cc||cc||cc||cc}
\hline
\hline
\multicolumn{9}{c}{BAO} \\
\hline
      & \multicolumn{2}{c||}{Multi-tracers} & \multicolumn{2}{c||}{LRG} & \multicolumn{2}{c||}{ELG} & \multicolumn{2}{c}{QSO} \\
\hline
  $z$ & $\sigma_{\Omega_m}$ & $\sigma_{\alpha}$ & $\sigma_{\Omega_m}$ & $\sigma_{\alpha}$ & $\sigma_{\Omega_m}$ & $\sigma_{\alpha}$ & $\sigma_{\Omega_m}$ & $\sigma_{\alpha}$  \\
\hline
\hline
0.3 & 1.301 & 11.894 &  1.529 & 15.218   &  1.415 & 13.589  & 66.832 & 625.066 \\
0.5 & 0.492 & 2.962 &   0.592 & 3.900   &   0.549 & 3.487   & 11.665 & 71.929 \\
0.7 & 0.278 & 1.317 &   0.363 & 1.879   &   0.321 & 1.603   & 4.867 & 23.517 \\
0.9 & 0.239 & 0.983 &   0.430 & 1.848   &   0.276 & 1.168   & 2.417 & 9.937 \\
1.1 & 0.304 & 1.128 &   1.797 & 6.662   &   0.335 & 1.248   & 1.457 & 5.346 \\
1.3 & 0.544 & 1.838 & $-$   & $-$       &   0.858 & 2.884   & 0.940 & 3.178 \\
1.5 & 0.756 & 2.403 & $-$   &  $-$   & $-$   &  $-$       &  0.756 & 2.403 \\
1.7 & 0.686 & 2.078 & $-$   &  $-$   & $-$   &  $-$       &  0.686 & 2.078 \\
1.9 & 0.638 & 1.865 & $-$   &  $-$   & $-$   &  $-$       &  0.638 & 1.865 \\
2.1 & 0.605 & 1.719 & $-$   &  $-$   & $-$   &  $-$       &  0.605 & 1.719 \\
2.3 & 0.585 & 1.626 & $-$   &  $-$   & $-$   &  $-$       &  0.585 & 1.626 \\
2.5 & 0.571 & 1.560 & $-$   &  $-$   & $-$   &  $-$       &  0.571 & 1.560 \\
2.7 & 0.564 & 1.520 & $-$   &  $-$   & $-$   &  $-$       &  0.564 & 1.520 \\
2.9 & 0.566 & 1.506 & $-$   &  $-$   & $-$   &  $-$       &  0.566 & 1.506 \\
3.1 & 0.570 & 1.505 & $-$   &  $-$   & $-$   &  $-$       &  0.570 & 1.505 \\
3.3 & 0.576 & 1.510 & $-$   &  $-$   & $-$   &  $-$       &  0.576 & 1.510 \\
3.5 & 0.588 & 1.531 & $-$   &  $-$   & $-$   &  $-$       &  0.588 & 1.531 \\
3.7 & 0.664 & 1.716 & $-$   &  $-$   & $-$   &  $-$       &  0.664 & 1.716 \\
3.9 & 0.755 & 1.939 & $-$   &  $-$   & $-$   &  $-$       &  0.755 & 1.939 \\
\hline
Total & 0.058 & 0.220   &   0.191 & 0.945   &   0.114 & 0.488   &   0.150 & 0.425 \\
\hline
\hline
\end{tabular}}
\end{minipage}
\end{center}
\end{table*}}}

{\renewcommand{\tabcolsep}{1.5mm}
{\renewcommand{\arraystretch}{1.8}
\begin{table*}
\begin{center}
\begin{minipage}{0.75\textwidth}
\caption{Errors for J-PAS $8500$ deg$^{2}$ using baryons and total matter. As we are focusing only on geometrical cosmological quantities, there is no difference between the baryons and the total matter scenario in this case.} \label{tab:results_JPAS8500-BAO}
\resizebox*{0.75\textwidth}{!}{
\begin{tabular}{c|cc||cc||cc||cc}
\hline
\hline
\multicolumn{9}{c}{BAO} \\
\hline
      & \multicolumn{2}{c||}{Multi-tracers} & \multicolumn{2}{c||}{LRG} & \multicolumn{2}{c||}{ELG} & \multicolumn{2}{c}{QSO} \\
\hline
  $z$ & $\sigma_{\Omega_m}$ & $\sigma_{\alpha}$ & $\sigma_{\Omega_m}$ & $\sigma_{\alpha}$ & $\sigma_{\Omega_m}$ & $\sigma_{\alpha}$ & $\sigma_{\Omega_m}$ & $\sigma_{\alpha}$  \\
\hline
\hline
0.3 & 0.892 & 8.159 &   1.049 & 10.439  &  0.971 & 9.322   & 45.846 & 428.792 \\
0.5 & 0.337 & 2.032 &   0.406 & 2.675   &  0.376 & 2.392   & 8.002 & 49.343 \\
0.7 & 0.191 & 0.903 &   0.249 & 1.289   &  0.220 & 1.100   & 3.338 & 16.133 \\
0.9 & 0.164 & 0.674 &   0.295 & 1.267   &  0.189 & 0.801   & 1.658 & 6.817 \\
1.1 & 0.208 & 0.774 &   1.232 & 4.570   &  0.230 & 0.856   & 0.999 & 3.668 \\
1.3 & 0.373 & 1.261 & $-$   & $-$       &  0.589 & 1.978   & 0.645 & 2.180  \\
1.5 & 0.519 & 1.649 & $-$   &  $-$   & $-$   &  $-$       &  0.519 & 1.649 \\
1.7 & 0.470 & 1.426 & $-$   &  $-$   & $-$   &  $-$       &  0.470 & 1.426 \\
1.9 & 0.438 & 1.279 & $-$   &  $-$   & $-$   &  $-$       &  0.438 & 1.279 \\
2.1 & 0.415 & 1.179 & $-$   &  $-$   & $-$   &  $-$       &  0.415 & 1.179 \\
2.3 & 0.402 & 1.116 & $-$   &  $-$   & $-$   &  $-$       &  0.585 & 1.626 \\
2.5 & 0.392 & 1.070 & $-$   &  $-$   & $-$   &  $-$       &  0.392 & 1.070 \\
2.7 & 0.387 & 1.042 & $-$   &  $-$   & $-$   &  $-$       &  0.387 & 1.042 \\
2.9 & 0.388 & 1.033 & $-$   &  $-$   & $-$   &  $-$       &  0.388 & 1.033 \\
3.1 & 0.391 & 1.033 & $-$   &  $-$   & $-$   &  $-$       &  0.391 & 1.033 \\
3.3 & 0.395 & 1.036 & $-$   &  $-$   & $-$   &  $-$       &  0.395 & 1.036 \\
3.5 & 0.403 & 1.050 & $-$   &  $-$   & $-$   &  $-$       &  0.403 & 1.050 \\
3.7 & 0.455 & 1.177 & $-$   &  $-$   & $-$   &  $-$       &  0.455 & 1.177 \\
3.9 & 0.518 & 1.330 & $-$   &  $-$   & $-$   &  $-$       &  0.518 & 1.330 \\
\hline
Total & 0.040 & 0.151   &   0.131 & 0.648   &   0.078 & 0.335   &   0.103 & 0.292 \\
\hline
\hline
\end{tabular}}
\end{minipage}
\end{center}
\end{table*}}}


\clearpage

\section{Tables for BAO + RSD}
\label{AppenC}

{\renewcommand{\tabcolsep}{1.5mm}
{\renewcommand{\arraystretch}{1.8}
\begin{table}[hbt!]
\begin{minipage}{0.5\columnwidth}
\caption{Errors for \textit{Euclid} using baryons and total matter.}\label{tab:results_EUCLID-BAO+RSD}
\begin{center}
\resizebox{0.6\columnwidth}{!}{
\begin{tabular}{c|cc|cc}
\hline
\hline
\multicolumn{5}{c}{BAO+RSD }\\
\hline
 & \multicolumn{4}{c}{ELG}\\
 & \multicolumn{2}{c}{tot. m.} & \multicolumn{2}{c}{bar. m.} \\
\hline
$z$ & $\sigma_{\Omega_m}$ & $\sigma_{\alpha}$ & $\sigma_{\Omega_m}$ & $\sigma_{\alpha}$ \\
\hline
\hline
1.00 & 0.040 & 0.119 &  0.10 & 0.404 \\
1.20 & 0.037 & 0.104 & 0.095 & 0.346 \\
1.40 & 0.038 & 0.099 & 0.093 & 0.312 \\
1.65 & 0.036 & 0.087 & 0.083 & 0.258 \\
\hline
Total & 0.019 & 0.087 & 0.041 & 0.138 \\
\hline
\hline
\end{tabular}}
\end{center}
\end{minipage}
\end{table}}}

{\renewcommand{\tabcolsep}{1.5mm}
{\renewcommand{\arraystretch}{1.8}
\begin{table}[hbt!]
\begin{minipage}{0.5\columnwidth}
\caption{Errors for DESI using baryons and total matter.}\label{tab:results_DESI-BAO+RSD}
\begin{center}
\resizebox{0.6\columnwidth}{!}{
\begin{tabular}{c|cc|cc}
\hline
\hline
\multicolumn{5}{c}{BAO+RSD }\\
\hline
      & \multicolumn{4}{c}{Multi-tracers} \\
      & \multicolumn{2}{c}{tot. m.} & \multicolumn{2}{c}{bar. m.} \\
\hline
  $z$ & $\sigma_{\Omega_m}$ & $\sigma_{\alpha}$  & $\sigma_{\Omega_m}$ & $\sigma_{\alpha}$ \\
\hline
\hline
0.1 & 1.013 & 4.683 & 1.289 & 3.369 \\
0.3 & 0.137 & 0.470 & 0.183 & 0.632 \\
0.5 & 0.155 & 0.511 & 0.201 & 1.069 \\
0.7 & 0.038 & 0.145 & 0.084 & 0.387 \\
0.9 & 0.034 & 0.114 & 0.083 & 0.338 \\
1.1 & 0.041 & 0.120 & 0.104 & 0.392 \\
1.3 & 0.040 & 0.126 & 0.076 & 0.384 \\
1.5 & 0.054 & 0.154 & 0.132 & 0.570 \\
1.7 & 0.100 & 0.252 & 0.267 & 1.019 \\
\hline
Total & 0.017 & 0.054 & 0.036 & 0.150 \\
\hline
\hline
\end{tabular}}
\end{center}
\end{minipage}
\end{table}}}

{\renewcommand{\tabcolsep}{1.5mm}
{\renewcommand{\arraystretch}{1.8}
\begin{table*}
\begin{center}
\begin{minipage}{0.95\textwidth}
\caption{Errors for J-PAS $4000$ deg$^{2}$ using baryons and total matter.} \label{tab:results_JPAS4000-BAO+RSD}
\resizebox*{\textwidth}{!}{
\begin{tabular}{c|cc|cc||cc|cc||cc|cc||cc|cc}
\hline
\hline
\multicolumn{17}{c}{BAO+RSD }\\
\hline
      & \multicolumn{4}{c||}{Multi-tracers} & \multicolumn{4}{c||}{LRG} & \multicolumn{4}{c||}{ELG} & \multicolumn{4}{c}{QSO} \\
      & \multicolumn{2}{c}{tot. m.} & \multicolumn{2}{c||}{bar. m.} & \multicolumn{2}{c}{tot. m.} & \multicolumn{2}{c||}{bar. m.} & \multicolumn{2}{c}{tot. m.} & \multicolumn{2}{c||}{bar. m.} & \multicolumn{2}{c}{tot. m.} & \multicolumn{2}{c}{bar. m.} \\
\hline
  $z$ & $\sigma_{\Omega_m}$ & $\sigma_{\alpha}$  & $\sigma_{\Omega_m}$ & $\sigma_{\alpha}$
      & $\sigma_{\Omega_m}$ & $\sigma_{\alpha}$  & $\sigma_{\Omega_m}$ & $\sigma_{\alpha}$
      & $\sigma_{\Omega_m}$ & $\sigma_{\alpha}$  & $\sigma_{\Omega_m}$ & $\sigma_{\alpha}$
      & $\sigma_{\Omega_m}$ & $\sigma_{\alpha}$  & $\sigma_{\Omega_m}$ & $\sigma_{\alpha}$ \\
\hline
\hline
0.3 & 0.124 & 0.753 & 0.185 & 1.030 &   0.286 & 0.894 & 0.369 & 1.198   &   0.208 & 0.836 & 0.291 & 1.126   &  6.692 & 26.666 & 9.464 & 36.166 \\
0.5 & 0.078 & 0.383 & 0.144 & 0.784 &   0.139 & 0.420 & 0.178 & 0.911   &   0.105 & 0.393 & 0.159 & 0.831   &  1.532 & 5.510  & 2.251 & 11.465 \\
0.7 & 0.064 & 0.253 & 0.144 & 0.668 &   0.105 & 0.301 & 0.176 & 0.928   &   0.079 & 0.266 & 0.155 & 0.769   &  0.856 & 2.754  & 1.590 & 7.691 \\
0.9 & 0.071 & 0.227 & 0.166 & 0.692 &   0.139 & 0.368 & 0.283 & 1.291   &   0.081 & 0.246 & 0.186 & 0.809   &  0.547 & 1.587  & 1.238 & 5.272 \\
1.1 & 0.099 & 0.279 & 0.238 & 0.913 &   0.565 & 1.375 & 1.307 & 5.107   &   0.107 & 0.300 & 0.260 & 0.998   &  0.406 & 1.081  & 1.001 & 3.820 \\
1.3 & 0.184 & 0.474 & 0.451 & 1.572 & $-$   & $-$    & $-$   &  $-$     &   0.269 & 0.701 & 0.687 & 2.378   &  0.310 & 0.772  & 0.769 & 2.693 \\
1.5 & 0.282 & 0.663 & 0.679 & 2.225 & $-$   & $-$    & $-$   &  $-$   & $-$   & $-$    & $-$   &  $-$       &  0.282 & 0.663  &  0.679 & 2.225  \\
1.7 & 0.280 & 0.626 & 0.649 & 2.013 & $-$   & $-$    & $-$   &  $-$   & $-$   & $-$    & $-$   &  $-$       &  0.280 & 0.626 & 0.649 & 2.013  \\
1.9 & 0.280 & 0.601 & 0.621 & 1.847 & $-$   & $-$    & $-$   &  $-$   & $-$   & $-$    & $-$   &  $-$       &  0.280 & 0.601 & 0.621 & 1.847 \\
2.1 & 0.281 & 0.581 & 0.599 & 1.718 & $-$   & $-$    & $-$   &  $-$   & $-$   & $-$    & $-$   &  $-$       &  0.281 & 0.581 & 0.599 & 1.718 \\
2.3 & 0.285 & 0.573 & 0.584 & 1.625 & $-$   & $-$    & $-$   &  $-$   & $-$   & $-$    & $-$   &  $-$       &  0.285 & 0.573 & 0.584 & 1.625 \\
2.5 & 0.288 & 0.566 & 0.571 & 1.548 & $-$   & $-$    & $-$   &  $-$   & $-$   & $-$    & $-$   &  $-$       &  0.288 & 0.566 & 0.571 & 1.548 \\
2.7 & 0.293 & 0.564 & 0.563 & 1.490 & $-$   & $-$    & $-$   &  $-$   & $-$   & $-$    & $-$   &  $-$       &  0.293 & 0.564 & 0.563 & 1.490 \\
2.9 & 0.301 & 0.569 & 0.562 & 1.456 & $-$   & $-$    & $-$   &  $-$   & $-$   & $-$    & $-$   &  $-$       &  0.301 & 0.569 & 0.562 & 1.456 \\
3.1 & 0.310 & 0.577 & 0.563 & 1.430 & $-$   & $-$    & $-$   &  $-$   & $-$   & $-$    & $-$   &  $-$       &  0.310 & 0.577 & 0.563 & 1.430 \\
3.3 & 0.319 & 0.586 & 0.564 & 1.408 & $-$   & $-$    & $-$   &  $-$   & $-$   & $-$    & $-$   &  $-$       &  0.319 & 0.586 & 0.564 & 1.408 \\
3.5 & 0.330 & 0.601 & 0.570 & 1.401 & $-$   & $-$    & $-$   &  $-$   & $-$   & $-$    & $-$   &  $-$       &  0.330 & 0.601 & 0.570 & 1.401 \\
3.7 & 0.377 & 0.681 & 0.639 & 1.545 & $-$   & $-$    & $-$   &  $-$   & $-$   & $-$    & $-$   &  $-$       &  0.377 & 0.681 & 0.639 & 1.545 \\
3.9 & 0.434 & 0.776 & 0.720 & 1.718 & $-$   & $-$    & $-$   &  $-$   & $-$   & $-$    & $-$   &  $-$       &  0.434 & 0.776 & 0.720 & 1.718 \\
\hline
Total & 0.030 & 0.082 & 0.049 & 0.176   &   0.068 & 0.189 & 0.104 & 0.499   &   0.043 & 0.132 & 0.081 & 0.345   &   0.077 & 0.154 & 0.134 & 0.370 \\
\hline
\hline
\end{tabular}}
\end{minipage}
\end{center}
\end{table*}}}

{\renewcommand{\tabcolsep}{1.5mm}
{\renewcommand{\arraystretch}{1.8}
\begin{table*}
\begin{center}
\begin{minipage}{0.95\textwidth}
\caption{Errors for J-PAS $8500$ deg$^{2}$ using baryons and total matter.} \label{tab:results_JPAS8500-BAO+RSD}
\resizebox*{\textwidth}{!}{
\begin{tabular}{c|cc|cc||cc|cc||cc|cc||cc|cc}
\hline
\hline
\multicolumn{17}{c}{BAO+RSD }\\
\hline
      & \multicolumn{4}{c||}{Multi-tracers} & \multicolumn{4}{c||}{LRG} & \multicolumn{4}{c||}{ELG} & \multicolumn{4}{c}{QSO} \\
      & \multicolumn{2}{c}{tot. m.} & \multicolumn{2}{c||}{bar. m.} & \multicolumn{2}{c}{tot. m.} & \multicolumn{2}{c||}{bar. m.} & \multicolumn{2}{c}{tot. m.} & \multicolumn{2}{c||}{bar. m.} & \multicolumn{2}{c}{tot. m.} & \multicolumn{2}{c}{bar. m.} \\
\hline
  $z$ & $\sigma_{\Omega_m}$ & $\sigma_{\alpha}$  & $\sigma_{\Omega_m}$ & $\sigma_{\alpha}$
      & $\sigma_{\Omega_m}$ & $\sigma_{\alpha}$  & $\sigma_{\Omega_m}$ & $\sigma_{\alpha}$
      & $\sigma_{\Omega_m}$ & $\sigma_{\alpha}$  & $\sigma_{\Omega_m}$ & $\sigma_{\alpha}$
      & $\sigma_{\Omega_m}$ & $\sigma_{\alpha}$  & $\sigma_{\Omega_m}$ & $\sigma_{\alpha}$ \\
\hline
\hline
0.3 & 0.085 & 0.516 & 0.127 & 0.706   &   0.196 & 0.613 & 0.253 & 0.822   &   0.143 & 0.574 & 0.200 & 0.772   &   4.590 & 18.293 & 6.492 & 24.810 \\
0.5 & 0.054 & 0.263 & 0.099 & 0.538   &   0.096 & 0.288 & 0.122 & 0.625   &   0.072 & 0.270 & 0.109 & 0.570   &   1.051 & 3.780 & 1.544 & 7.865 \\
0.7 & 0.044 & 0.174 & 0.099 & 0.458   &   0.072 & 0.206 & 0.121 & 0.637   &   0.054 & 0.182 & 0.106 & 0.527   &   0.587 & 1.889 & 1.091 & 5.276  \\
0.9 & 0.049 & 0.156 & 0.114 & 0.475   &   0.096 & 0.252 & 0.194 & 0.886   &   0.055 & 0.169 & 0.128 & 0.555   &   0.375 & 1.089 & 0.849 & 3.617 \\
1.1 & 0.068 & 0.191 & 0.164 & 0.626   &   0.388 & 0.943 & 0.896 & 3.503   &   0.073 & 0.206 & 0.178 & 0.685   &   0.278 & 0.742 & 0.687 & 2.621 \\
1.3 & 0.126 & 0.325 & 0.309 & 1.078   & $-$   & $-$   & $-$   &  $-$      &   0.184 & 0.481 & 0.471 & 1.632   &   0.213 & 0.530 & 0.528 & 1.848 \\
1.5 & 0.194 & 0.454 & 0.466 & 1.526   & $-$   & $-$    & $-$   &  $-$   & $-$   & $-$    & $-$   &  $-$        &   0.194 & 0.454 & 0.466 & 1.526 \\
1.7 & 0.192 & 0.430 & 0.445 & 1.381   & $-$   & $-$    & $-$   &  $-$   & $-$   & $-$    & $-$   &  $-$        &   0.192 & 0.430 & 0.445 & 1.381 \\
1.9 & 0.192 & 0.412 & 0.426 & 1.267   & $-$   & $-$    & $-$   &  $-$   & $-$   & $-$    & $-$   &  $-$        &   0.192 & 0.412 & 0.426 & 1.267 \\
2.1 & 0.193 & 0.399 & 0.411 & 1.179   & $-$   & $-$    & $-$   &  $-$   & $-$   & $-$    & $-$   &  $-$        &   0.193 & 0.399 & 0.411 & 1.179 \\
2.3 & 0.195 & 0.393 & 0.401 & 1.114   & $-$   & $-$    & $-$   &  $-$   & $-$   & $-$    & $-$   &  $-$        &   0.195 & 0.393 & 0.401 & 1.114 \\
2.5 & 0.198 & 0.388 & 0.392 & 1.062   & $-$   & $-$    & $-$   &  $-$   & $-$   & $-$    & $-$   &  $-$        &   0.198 & 0.388 & 0.392 & 1.062 \\
2.7 & 0.201 & 0.387 & 0.386 & 1.022   & $-$   & $-$    & $-$   &  $-$   & $-$   & $-$    & $-$   &  $-$        &   0.201 & 0.387 & 0.386 & 1.022 \\
2.9 & 0.207 & 0.390 & 0.386 & 0.999   & $-$   & $-$    & $-$   &  $-$   & $-$   & $-$    & $-$   &  $-$        &   0.207 & 0.390 & 0.386 & 0.999 \\
3.1 & 0.213 & 0.396 & 0.386 & 0.981   & $-$   & $-$    & $-$   &  $-$   & $-$   & $-$    & $-$   &  $-$        &   0.213 & 0.396 & 0.386 & 0.981 \\
3.3 & 0.219 & 0.402 & 0.387 & 0.966   & $-$   & $-$    & $-$   &  $-$   & $-$   & $-$    & $-$   &  $-$        &   0.219 & 0.402 & 0.387 & 0.966 \\
3.5 & 0.227 & 0.412 & 0.391 & 0.961   & $-$   & $-$    & $-$   &  $-$   & $-$   & $-$    & $-$   &  $-$        &   0.227 & 0.412 & 0.391 & 0.961 \\
3.7 & 0.259 & 0.467 & 0.438 & 1.060   & $-$   & $-$    & $-$   &  $-$   & $-$   & $-$    & $-$   &  $-$        &   0.259 & 0.467 & 0.438 & 1.060 \\
3.9 & 0.298 & 0.533 & 0.494 & 1.178   & $-$   & $-$    & $-$   &  $-$   & $-$   & $-$    & $-$   &  $-$        &   0.298 & 0.533 & 0.494 & 1.178 \\
\hline
Total & 0.021 & 0.056 & 0.034 & 0.121   &   0.046 & 0.129 & 0.071 & 0.342   &    0.030 & 0.090 & 0.055 & 0.236   &    0.053 & 0.106 & 0.092 & 0.254 \\
\hline
\hline
\end{tabular}}
\end{minipage}
\end{center}
\end{table*}}}


\clearpage

\section{Tables for BAO + RSD + PS}
\label{AppenD}

{\renewcommand{\tabcolsep}{1.5mm}
{\renewcommand{\arraystretch}{1.8}
\begin{table}[hbt!]
\begin{minipage}{0.5\columnwidth}
\caption{Errors for \textit{Euclid} using baryons and total matter.}\label{tab:results_EUCLID-FULL}
\begin{center}
\resizebox*{0.6\columnwidth}{!}{
\begin{tabular}{c|cc|cc}
\hline
\hline
\multicolumn{5}{c}{BAO+RSD+PS}\\
\hline
 & \multicolumn{4}{c}{ELG}\\
 & \multicolumn{2}{c}{tot. m.} & \multicolumn{2}{c}{bar. m.} \\
\hline
$z$ & $\sigma_{\Omega_m}$ & $\sigma_{\alpha}$ & $\sigma_{\Omega_m}$ & $\sigma_{\alpha}$ \\
\hline
\hline
1.00 & 0.0029 & 0.018 & 0.0023 & 0.038 \\
1.20 & 0.0027 & 0.014 & 0.0023 & 0.032 \\
1.40 & 0.0027 & 0.013 & 0.0025 & 0.028 \\
1.65 & 0.0025 & 0.011 & 0.0024 & 0.022 \\
\hline
Total & 0.001 & 0.007 & 0.001 & 0.014 \\
\hline
\hline
\end{tabular}}
\end{center}
\end{minipage}
\end{table}}}

{\renewcommand{\tabcolsep}{1.5mm}
{\renewcommand{\arraystretch}{1.8}
\begin{table}[hbt!]
\begin{minipage}{0.5\columnwidth}
\caption{Errors for DESI using baryons and total matter.}\label{tab:results_DESI-FULL}
\begin{center}
\resizebox*{0.6\columnwidth}{!}{
\begin{tabular}{c|cc|cc}
\hline
\hline
\multicolumn{5}{c}{BAO+RSD+PS}\\
\hline
      & \multicolumn{4}{c}{Multi-tracers} \\
      & \multicolumn{2}{c}{tot. m.} & \multicolumn{2}{c}{bar. m.} \\
\hline
  $z$ & $\sigma_{\Omega_m}$ & $\sigma_{\alpha}$  & $\sigma_{\Omega_m}$ & $\sigma_{\alpha}$ \\
\hline
\hline
0.1 & 0.014 & 0.354 & 0.014 & 0.194 \\
0.3 & 0.006 & 0.102 & 0.005 & 0.084 \\
0.5 & 0.008 & 0.101 & 0.007 & 0.127 \\
0.7 & 0.003 & 0.027 & 0.002 & 0.040 \\
0.9 & 0.003 & 0.019 & 0.002 & 0.034 \\
1.1 & 0.003 & 0.017 & 0.002 & 0.036 \\
1.3 & 0.003 & 0.021 & 0.002 & 0.039 \\
1.5 & 0.004 & 0.023 & 0.003 & 0.057 \\
1.7 & 0.006 & 0.036 & 0.007 & 0.101 \\
\hline
Total & 0.001 & 0.009 & 0.001 & 0.016 \\
\hline
\hline
\end{tabular}}
\end{center}
\end{minipage}
\end{table}}}

{\renewcommand{\tabcolsep}{1.5mm}
{\renewcommand{\arraystretch}{1.8}
\begin{table*}
\begin{center}
\begin{minipage}{0.925\textwidth}
\caption{Errors for J-PAS $4000$ deg$^{2}$ using baryons and total matter.} \label{tab:results_JPAS4000-FULL}
\resizebox*{\textwidth}{!}{
\begin{tabular}{c|cc|cc||cc|cc||cc|cc||cc|cc}
\hline
\hline
\multicolumn{17}{c}{BAO+RSD+PS}\\
\hline
      & \multicolumn{4}{c||}{Multi-tracers} & \multicolumn{4}{c||}{LRG} & \multicolumn{4}{c||}{ELG} & \multicolumn{4}{c}{QSO} \\
      & \multicolumn{2}{c}{tot. m.} & \multicolumn{2}{c||}{bar. m.} & \multicolumn{2}{c}{tot. m.} & \multicolumn{2}{c||}{bar. m.} & \multicolumn{2}{c}{tot. m.} & \multicolumn{2}{c||}{bar. m.} & \multicolumn{2}{c}{tot. m.} & \multicolumn{2}{c}{bar. m.} \\
\hline
  $z$ & $\sigma_{\Omega_m}$ & $\sigma_{\alpha}$  & $\sigma_{\Omega_m}$ & $\sigma_{\alpha}$
      & $\sigma_{\Omega_m}$ & $\sigma_{\alpha}$  & $\sigma_{\Omega_m}$ & $\sigma_{\alpha}$
      & $\sigma_{\Omega_m}$ & $\sigma_{\alpha}$  & $\sigma_{\Omega_m}$ & $\sigma_{\alpha}$
      & $\sigma_{\Omega_m}$ & $\sigma_{\alpha}$  & $\sigma_{\Omega_m}$ & $\sigma_{\alpha}$ \\
\hline
\hline
0.3 & 0.010 & 0.141 & 0.009 & 0.099 &   0.011 & 0.219 & 0.010 & 0.188   &   0.011 & 0.156 & 0.010 & 0.118   &  0.272 & 3.801  & 0.256 & 2.844 \\
0.5 & 0.006 & 0.073 & 0.006 & 0.069 &   0.007 & 0.103 & 0.007 & 0.147   &   0.007 & 0.077 & 0.006 & 0.083   &  0.080 & 0.870  & 0.071 & 0.918 \\
0.7 & 0.005 & 0.047 & 0.004 & 0.060 &   0.006 & 0.066 & 0.006 & 0.139   &   0.006 & 0.049 & 0.005 & 0.071   &  0.047 & 0.421  & 0.041 & 0.622 \\
0.9 & 0.005 & 0.039 & 0.004 & 0.068 &   0.007 & 0.070 & 0.009 & 0.166   &   0.006 & 0.040 & 0.005 & 0.074   &  0.030 & 0.234  & 0.027 & 0.460 \\
1.1 & 0.007 & 0.041 & 0.006 & 0.086 &   0.026 & 0.245 & 0.033 & 0.589   &   0.007 & 0.043 & 0.006 & 0.090   &  0.022 & 0.155  & 0.023 & 0.358 \\
1.3 & 0.011 & 0.063 & 0.011 & 0.144 & $-$   & $-$    & $-$   &  $-$     &   0.016 & 0.088 & 0.015 & 0.201   &  0.017 & 0.109  & 0.019 & 0.256 \\
1.5 & 0.015 & 0.093 & 0.017 & 0.205 & $-$   & $-$    & $-$   &  $-$   & $-$   & $-$    & $-$   &  $-$       &  0.015 & 0.093  & 0.017 & 0.205 \\
1.7 & 0.014 & 0.088 & 0.017 & 0.179 & $-$   & $-$    & $-$   &  $-$   & $-$   & $-$    & $-$   &  $-$       &  0.014 & 0.088  & 0.017 & 0.179 \\
1.9 & 0.014 & 0.086 & 0.016 & 0.159 & $-$   & $-$    & $-$   &  $-$   & $-$   & $-$    & $-$   &  $-$       &  0.014 & 0.086  & 0.016 & 0.159 \\
2.1 & 0.014 & 0.084 & 0.016 & 0.145 & $-$   & $-$    & $-$   &  $-$   & $-$   & $-$    & $-$   &  $-$       &  0.014 & 0.084  & 0.016 & 0.145 \\
2.3 & 0.014 & 0.084 & 0.016 & 0.135 & $-$   & $-$    & $-$   &  $-$   & $-$   & $-$    & $-$   &  $-$       &  0.014 & 0.084  & 0.016 & 0.135 \\
2.5 & 0.014 & 0.085 & 0.016 & 0.128 & $-$   & $-$    & $-$   &  $-$   & $-$   & $-$    & $-$   &  $-$       &  0.014 & 0.085  & 0.016 & 0.128 \\
2.7 & 0.014 & 0.086 & 0.016 & 0.124 & $-$   & $-$    & $-$   &  $-$   & $-$   & $-$    & $-$   &  $-$       &  0.014 & 0.086  & 0.016 & 0.124 \\
2.9 & 0.015 & 0.087 & 0.016 & 0.121 & $-$   & $-$    & $-$   &  $-$   & $-$   & $-$    & $-$   &  $-$       &  0.015 & 0.087  & 0.016 & 0.121 \\
3.1 & 0.015 & 0.090 & 0.016 & 0.120 & $-$   & $-$    & $-$   &  $-$   & $-$   & $-$    & $-$   &  $-$       &  0.015 & 0.090  & 0.016 & 0.120 \\
3.3 & 0.015 & 0.092 & 0.017 & 0.119 & $-$   & $-$    & $-$   &  $-$   & $-$   & $-$    & $-$   &  $-$       &  0.015 & 0.092  & 0.017 & 0.119 \\
3.5 & 0.016 & 0.095 & 0.017 & 0.120 & $-$   & $-$    & $-$   &  $-$   & $-$   & $-$    & $-$   &  $-$       &  0.016 & 0.095  & 0.017 & 0.120 \\
3.7 & 0.018 & 0.108 & 0.019 & 0.134 & $-$   & $-$    & $-$   &  $-$   & $-$   & $-$    & $-$   &  $-$       &  0.018 & 0.108  & 0.019 & 0.134 \\
3.9 & 0.020 & 0.124 & 0.022 & 0.150 & $-$   & $-$    & $-$   &  $-$   & $-$   & $-$    & $-$   &  $-$       &  0.020 & 0.124  & 0.022 & 0.150 \\
\hline
Total & 0.002 & 0.016 & 0.004 & 0.038   &   0.004 & 0.042 & 0.004 & 0.075   &   0.003 & 0.023 & 0.002 & 0.035   &   0.004 & 0.024 & 0.004 & 0.037 \\
\hline
\hline
\end{tabular}}
\end{minipage}
\end{center}
\end{table*}}}

{\renewcommand{\tabcolsep}{1.5mm}
{\renewcommand{\arraystretch}{1.8}
\begin{table*}
\begin{center}
\begin{minipage}{0.925\textwidth}
\caption{Errors for J-PAS $8500$ deg$^{2}$ using baryons and total matter.} \label{tab:results_JPAS8500-FULL}
\resizebox*{\textwidth}{!}{
\begin{tabular}{c|cc|cc||cc|cc||cc|cc||cc|cc}
\hline
\hline
\multicolumn{17}{c}{BAO+RSD+PS}\\
\hline
      & \multicolumn{4}{c||}{Multi-tracers} & \multicolumn{4}{c||}{LRG} & \multicolumn{4}{c||}{ELG} & \multicolumn{4}{c}{QSO} \\
      & \multicolumn{2}{c}{tot. m.} & \multicolumn{2}{c||}{bar. m.} & \multicolumn{2}{c}{tot. m.} & \multicolumn{2}{c||}{bar. m.} & \multicolumn{2}{c}{tot. m.} & \multicolumn{2}{c||}{bar. m.} & \multicolumn{2}{c}{tot. m.} & \multicolumn{2}{c}{bar. m.} \\
\hline
  $z$ & $\sigma_{\Omega_m}$ & $\sigma_{\alpha}$  & $\sigma_{\Omega_m}$ & $\sigma_{\alpha}$
      & $\sigma_{\Omega_m}$ & $\sigma_{\alpha}$  & $\sigma_{\Omega_m}$ & $\sigma_{\alpha}$
      & $\sigma_{\Omega_m}$ & $\sigma_{\alpha}$  & $\sigma_{\Omega_m}$ & $\sigma_{\alpha}$
      & $\sigma_{\Omega_m}$ & $\sigma_{\alpha}$  & $\sigma_{\Omega_m}$ & $\sigma_{\alpha}$ \\
\hline
\hline
0.3 & 0.007 & 0.097 & 0.006 & 0.068   &   0.007 & 0.151 & 0.007 & 0.129   &   0.007 & 0.107 & 0.007 & 0.081   &   0.186 & 2.608 & 0.176 & 1.951 \\
0.5 & 0.004 & 0.050 & 0.004 & 0.047   &   0.005 & 0.071 & 0.005 & 0.101   &   0.005 & 0.053 & 0.004 & 0.057   &   0.055 & 0.597 & 0.049 & 0.630 \\
0.7 & 0.004 & 0.032 & 0.003 & 0.041   &   0.004 & 0.045 & 0.004 & 0.095   &   0.004 & 0.033 & 0.003 & 0.049   &   0.032 & 0.289 & 0.028 & 0.427 \\
0.9 & 0.004 & 0.026 & 0.003 & 0.046   &   0.005 & 0.048 & 0.006 & 0.114   &   0.004 & 0.027 & 0.003 & 0.050   &   0.020 & 0.161 & 0.019 & 0.315 \\
1.1 & 0.005 & 0.028 & 0.004 & 0.059   &   0.018 & 0.168 & 0.022 & 0.404   &   0.005 & 0.029 & 0.004 & 0.062   &   0.015 & 0.106 & 0.015 & 0.246 \\
1.3 & 0.007 & 0.043 & 0.007 & 0.099   & $-$   & $-$   & $-$   &  $-$      &   0.011 & 0.060 & 0.010 & 0.138   &   0.011 & 0.074 & 0.013 & 0.176 \\
1.5 & 0.010 & 0.064 & 0.012 & 0.141   & $-$   & $-$    & $-$   &  $-$   & $-$   & $-$    & $-$   &  $-$       &   0.010 & 0.064 & 0.012 & 0.141 \\
1.7 & 0.010 & 0.061 & 0.012 & 0.123   & $-$   & $-$    & $-$   &  $-$   & $-$   & $-$    & $-$   &  $-$       &   0.010 & 0.061 & 0.012 & 0.123 \\
1.9 & 0.010 & 0.059 & 0.011 & 0.109   & $-$   & $-$    & $-$   &  $-$   & $-$   & $-$    & $-$   &  $-$       &   0.010 & 0.059 & 0.011 & 0.109 \\
2.1 & 0.010 & 0.058 & 0.011 & 0.099   & $-$   & $-$    & $-$   &  $-$   & $-$   & $-$    & $-$   &  $-$       &   0.010 & 0.058 & 0.011 & 0.099 \\
2.3 & 0.010 & 0.058 & 0.011 & 0.093   & $-$   & $-$    & $-$   &  $-$   & $-$   & $-$    & $-$   &  $-$       &   0.010 & 0.058 & 0.011 & 0.093 \\
2.5 & 0.010 & 0.058 & 0.011 & 0.088   & $-$   & $-$    & $-$   &  $-$   & $-$   & $-$    & $-$   &  $-$       &   0.010 & 0.058 & 0.011 & 0.088 \\
2.7 & 0.010 & 0.059 & 0.011 & 0.085   & $-$   & $-$    & $-$   &  $-$   & $-$   & $-$    & $-$   &  $-$       &   0.010 & 0.059 & 0.011 & 0.085 \\
2.9 & 0.010 & 0.060 & 0.011 & 0.083   & $-$   & $-$    & $-$   &  $-$   & $-$   & $-$    & $-$   &  $-$       &   0.010 & 0.060 & 0.011 & 0.083 \\
3.1 & 0.010 & 0.062 & 0.011 & 0.082   & $-$   & $-$    & $-$   &  $-$   & $-$   & $-$    & $-$   &  $-$       &   0.010 & 0.062 & 0.011 & 0.082 \\
3.3 & 0.010 & 0.063 & 0.011 & 0.082   & $-$   & $-$    & $-$   &  $-$   & $-$   & $-$    & $-$   &  $-$       &   0.010 & 0.063 & 0.011 & 0.082 \\
3.5 & 0.011 & 0.065 & 0.012 & 0.082   & $-$   & $-$    & $-$   &  $-$   & $-$   & $-$    & $-$   &  $-$       &   0.011 & 0.065 & 0.012 & 0.082 \\
3.7 & 0.012 & 0.074 & 0.013 & 0.092   & $-$   & $-$    & $-$   &  $-$   & $-$   & $-$    & $-$   &  $-$       &   0.012 & 0.074 & 0.013 & 0.092 \\
3.9 & 0.014 & 0.085 & 0.015 & 0.103   & $-$   & $-$    & $-$   &  $-$   & $-$   & $-$    & $-$   &  $-$       &   0.014 & 0.085 & 0.015 & 0.103 \\
\hline
Total & 0.001 & 0.011 & 0.001 & 0.013   &   0.003 & 0.029 & 0.003 & 0.051   &    0.002 & 0.015 & 0.002 & 0.024   &    0.003 & 0.017 & 0.003 & 0.025 \\
\hline
\hline
\end{tabular}}
\end{minipage}
\end{center}
\end{table*}}}




\label{lastpage}

\end{document}